%% file: manuscript.tex
\documentclass[prd,twocolumn,floatfix,nofootinbib]{revtex4}
\usepackage{amssymb}
\usepackage{amsmath}
\usepackage{graphicx,subfigure,color,dcolumn,booktabs,bm}
\usepackage{longtable,lscape}
\usepackage{txfonts}
\usepackage{overpic}
\usepackage{indentfirst}
\usepackage{cases}
\usepackage{multirow}
\usepackage{ulem}
\usepackage{enumerate}
\usepackage[colorlinks,
            citecolor=blue,
            anchorcolor=red,
            menucolor=red,
            linkcolor=red,
            filecolor=red,
            runcolor=red,
            urlcolor=blue,
            frenchlinks=false]{hyperref}

\allowdisplaybreaks

\begin{document}

\title{Truncation of the Radial Ladder in Heavy Quarkonia}
\author{Wen-Xuan Zhang$^{1,2,3,4}$}
\email{zhangwx89@outlook.com}
\author{Si-Qiang Luo$^{1,2,3,4}$}
\email{luosq15@lzu.edu.cn}
\author{Xiang Liu$^{1,2,3,4}$}
\email{xiangliu@lzu.edu.cn}
\affiliation{$^1$School of Physical Science and Technology, Lanzhou University, Lanzhou 730000, China \\
$^2$Lanzhou Center for Theoretical Physics, Key Laboratory of Theoretical Physics of Gansu Province,\\
Key Laboratory of Quantum Theory and Applications of MoE,\\
Gansu Provincial Research Center for Basic Disciplines of Quantum Physics, Lanzhou University, Lanzhou 730000, China\\
$^3$MoE Frontiers Science Center for Rare Isotopes, Lanzhou University, Lanzhou 730000, China\\
$^4$Research Center for Hadron and CSR Physics, Lanzhou University and Institute of Modern Physics of CAS, Lanzhou 730000, China}

\begin{abstract}
A fundamental open question in hadron spectroscopy is whether the radial excitation ladder of quarkonia truncates at a finite level---a possibility that would challenge the conventional quark-antiquark bound-state picture and offer decisive clues to the nonperturbative strong interaction. Taking advantage of the newly observed high-mass hadronic states, we address this issue by solving a screened Godfrey--Isgur Hamiltonian for charmonium and bottomonium using the Gaussian expansion method. The calculated spectra saturate at $4.74\,\mathrm{GeV}$ ($c\bar{c}$) and $11.67\,\mathrm{GeV}$ ($b\bar{b}$), while root-mean-square radii grow to $\sim10\,\mathrm{fm}$---an order of magnitude above the confinement scale---where adjacent mass gaps drop below $10\,\mathrm{MeV}$. Combining a threshold-based mass-gap criterion, defined by the onset of mass--radius decoupling, with additional diagnostics, we locate the operational upper radial limits at $n\approx8$--$10$ for charmonium and $n\approx12$--$13$ for bottomonium. Beyond these limits, the conventional $q\bar{q}$ description ceases to apply. This work provides the first quantitative determination of these upper limits, and the proposed criterion is directly testable with forthcoming high-statistics data from BESIII, Belle II, and LHCb.
\end{abstract}

\maketitle

A quantitative description of color confinement, central to the nonperturbative strong interaction, remains an important issue in particle physics. Studying hadron spectroscopy from three perspectives---phenomenology, experiment, and lattice QCD---provides an effective approach to deepening our understanding of the nonperturbative QCD.

Inspired by the abundant charmonium discoveries in the 1970s \cite{E598:1974sol,SLAC-SP-017:1974ind,Abrams:1974yy,Tanenbaum:1975ef,Whitaker:1976hb,Siegrist:1976br,Rapidis:1977cv,Biddick:1977sv,Goldhaber:1977qn,DASP:1978dns}, the Cornell potential, $V(r)=a r+b/r$ \cite{Eichten:1974af, Eichten:1978tg, Eichten:1979ms}, successfully described low-lying hadron spectroscopy---a key achievement of the late twentieth century. However, since 2003, an increasing number of new hadronic states and related novel phenomena have been observed \cite{Liu:2013waa, Lebed:2016hpi, Richard:2016eis, Hosaka:2016pey, Chen:2016qju, Olsen:2017bmm, Guo:2017jvc, Brambilla:2019esw, Liu:2019zoy, Meng:2022ozq, Chen:2022asf, Liu:2024uxn, Wang:2025dur, Wang:2025sic, Bai:2026atm}, indicating that the quenched approximation is no longer adequate. Consequently, unquenched effects should be included when studying high-lying hadrons.

Taking mesons in the hadron family as an example, their mass spectra differ significantly between quenched and unquenched potential models \cite{Bai:2026atm}. The Cornell potential, as a typical quenched model, implies an unbounded meson mass spectrum, since the strong interaction is described by its linear term at long distances. However, unquenched effects modify the long-distance behavior of the strong-interaction potential, leading to meson energy levels that differ from those obtained in the quenched approximation \cite{Ding:1993uy,Li:2009nr,Song:2019screened,Chen:2024unquenched}. This suggests that the meson spectrum is truncated. A quantitative determination of this truncation is therefore both important and timely, as relevant studies are currently absent.

\begin{figure}[!htbp]
    \centering
    \includegraphics[width=0.98\columnwidth]{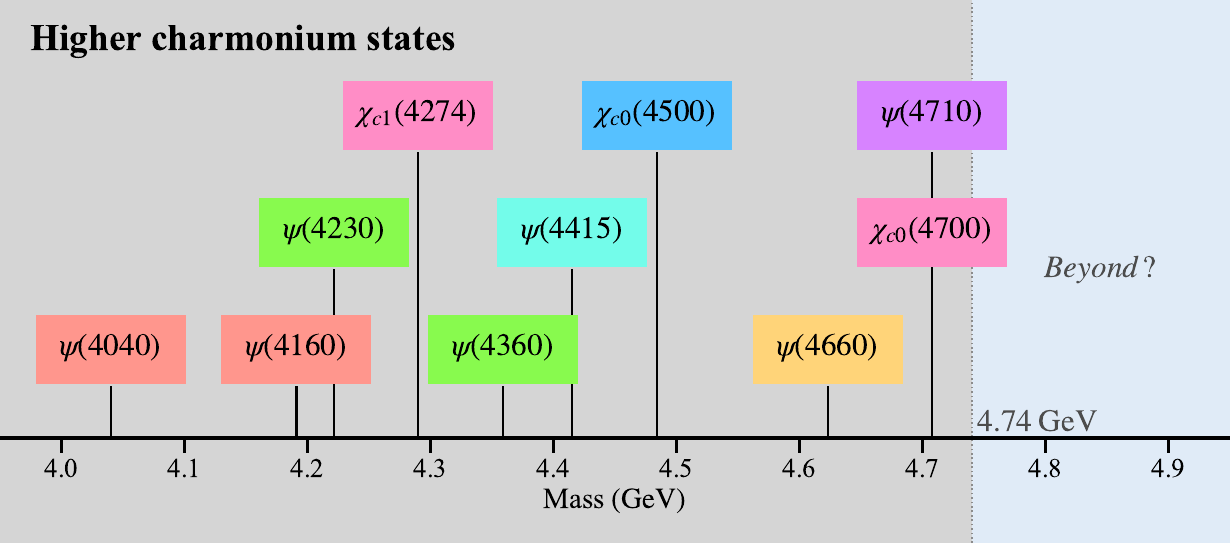}
    \caption{Some established charmonium states with masses below the calculated limiting mass of $4.74\,\mathrm{GeV}$.
    The question mark emphasizes the open issue of whether any additional charmonium states exist beyond this boundary.}
    \label{fig:charmonium-mass-axis}
\end{figure}

In experiments, the region above open-flavor thresholds has been extensively explored over the past two decades. The BaBar Collaboration first reported a broad structure near \(4.26\,\mathrm{GeV}\) in the \(\pi^+\pi^-J/\psi\) mass spectrum \cite{BaBar:2005hhc}, followed by the observation of the \(\psi(4260)\) (also referred to as \(\psi(4230)\)) in the \(\psi(2S)\pi^+\pi^-\) channel \cite{BaBar:2006ait} and the \(\psi(4660)\) \cite{Belle:2007umv}. In the bottomonium sector, Belle has observed the \(\Upsilon(10753)\), \(\Upsilon(10860)\), and \(\Upsilon(11020)\) \cite{Belle:2011aa}. More recently, high-statistics BESIII data have revealed rich line shapes and multiple resonant structures in channels such as \(e^+e^-\to K^+K^-J/\psi\), including the vector charmonium-like states \(\psi(4230)\), \(\psi(4500)\), and \(\psi(4710)\) near \(4.71\,\mathrm{GeV}\)---among the heaviest charmonium-like states observed to date \cite{BESIII:2023wqy,Li:2026jzk}.
These findings indicate that charmonium and bottomonium systems serve as ideal platforms for investigating the spectroscopic properties. In Figure~\ref{fig:charmonium-mass-axis}, we present the higher-lying observed charmonium states with masses $4.0$--$4.7\,\mathrm{GeV}$.

In this work, we address this issue by solving the screened, relativized Godfrey--Isgur (GI) Hamiltonian \cite{Godfrey:1985xj} for both charmonium and bottomonium with the Gaussian expansion method (GEM) \cite{Hiyama:2003cu,Hiyama:2012sma}, with parameters fixed by a staged fit to established heavy-meson states. We compute masses and root-mean-square (RMS) radii up to radial quantum number \(n=14\). The spectra are found to truncate at limiting masses of \(4.74\,\mathrm{GeV}\) for \(c\bar{c}\) and \(11.67\,\mathrm{GeV}\) for \(b\bar{b}\), while the RMS radii continue to diverge, reaching \(\sim 10\,\mathrm{fm}\)---an order of magnitude above the confinement scale. This mass--radius decoupling provides a physical signature of the breakdown of the bound-state picture. We locate the operational upper radial limits using a threshold-based mass-gap criterion, with additional diagnostics, yielding \(n\approx 8\)--\(10\) for charmonium and \(n\approx 12\)--\(13\) for bottomonium. Beyond these limits, the conventional \(q\bar{q}\) description ceases to apply, accounting for the absence of additional charmonium states above limiting masses. The proposed criterion is directly testable with forthcoming high-statistics data from BESIII, Belle II, and LHCb.

The model Hamiltonian is the relativized screened GI form
\begin{align}
    H &= \sum_i\sqrt{p_i^2+m_i^2} \nonumber \\
    &+ \sum_{i<j}\bigl(V_{ij}^{\rm Coul}+V_{ij}^{\rm conf}+V_{ij}^{\rm cont}+V_{ij}^{\rm so(\nu)}+V_{ij}^{\rm so(s)}+V_{ij}^{\rm tens}\bigr),
\end{align}
in which all interactions are folded with a Gaussian smearing kernel. The central modification is the screened confinement \cite{Song:2015nia,Wang:2019mhs,Wang:2020prx}
\begin{equation}
    V^{\rm conf}(r) = \int d^3\mathbf{r^{\prime}} \big(\frac{\sigma^{3}}{\pi^{3/2}}
    e^{-\sigma^2 (\mathbf{r}-\mathbf{r^{\prime}})^2}\big) \big(\frac{b(1-e^{-\mu r^{\prime}})}{\mu}+c\big),
\end{equation}
which asymptotes to the finite constant $b/\mu+c$. Orbital wave functions are expanded in the Gaussian basis
\begin{equation}
    \phi_{nl}(\mathbf{r}) = N_{nl}\, r^l\, e^{-\nu_n r^2} Y_{lm}(\mathbf{r}),
\end{equation}
with range parameters chosen in geometric progression
\begin{equation}
    \nu_n = r_n^{-2},\qquad r_n = r_{\rm min}\,a^{n-1},\qquad
    a = \Bigl(\frac{r_{\rm max}}{r_{\rm min}}\Bigr)^{1/(N_{\rm max}-1)}.
\end{equation}
With $N_{\rm max}=20$, $r_{\rm min}=0.1\,\mathrm{fm}$ and $r_{\rm max}=30\,\mathrm{fm}$, 
the same basis simultaneously describes the short-distance Coulomb interaction and the long-range multi-node structure. 
Because the basis functions are non-orthogonal, the variational problem is solved by a standard eigenvalue equation after random-matrix orthogonalization, 
yielding stable nodal wave functions. Algorithms and implementations are presented in the Supplemental Material.

The model parameters are determined by a staged $\chi^2$ fit with \texttt{ROOT::Minuit2} \cite{ROOT_NIMA_1997}.
In the first stage, system-independent quantities---constituent masses, the constant $c$, and the smearing parameters $\sigma_0$ and $s$---are optimized on the full set of established heavy mesons ($c\bar{c}$, $b\bar{b}$, $B_c$, $D_s$, $B_s$, $D$, and $B$).
Multiple initial conditions are tested and the solution with minimal $\chi^2$ is selected as the global optimum of this stage.
With those parameters fixed, the string tension $b$, the screening parameter $\mu$, and the relativistic exponents are then refitted separately in the charmonium and bottomonium sectors.
The second stage has a well-conditioned minimum, in which repeated runs from the starting point converge to the same parameter set.
This sector separation is required by the energy-scale dependence of the strong interaction and color confinement. The resulting parameters are listed in Table~\ref{tab:params}.

Alternative strategies fail physical checks of the limiting mass $M_\infty$ and the screening length $\mu^{-1}$:
\begin{enumerate}[(i)]
\item A fully unified fit with identical $b$ and $\mu$ for both sectors yields $M_\infty=4.83\,\mathrm{GeV}$ ($c\bar{c}$) and $11.52\,\mathrm{GeV}$ ($b\bar{b}$) with a common $\mu^{-1}\approx1.5\,\mathrm{fm}$, missing the expected sector dependence of the long-range dynamics.
\item Fitting charmonium alone without the first stage gives $M_\infty\approx4.61\,\mathrm{GeV}$ and $\mu^{-1}\approx1.25\,\mathrm{fm}$: the screening length is only marginally acceptable, yet the limiting mass remains unphysically low.
\item If the parameters $c$, $\sigma_0$, and $s$ are left free in the second stage, charmonium collapses to $M_\infty\approx4.62\,\mathrm{GeV}$ with $\mu^{-1}\approx1.19\,\mathrm{fm}$, still outside the physically preferred range.
\end{enumerate}
By contrast, the final solution produces $M_\infty=4.74\,\mathrm{GeV}$ ($c\bar{c}$) and $11.67\,\mathrm{GeV}$ ($b\bar{b}$), with $\mu^{-1}\approx1.4\,\mathrm{fm}$ and $\approx1.6\,\mathrm{fm}$, respectively.
The charmonium screening length lies within the lattice range $1.2$--$1.4\,\mathrm{fm}$ \cite{Bali:2005fu,Castorina:2007eb,Bulava:2019iut,Jiang:2023lmj,Kou:2024dml}, while bottomonium is longer, as expected for a heavier, more compact system.
Because the staged procedure is constrained by global minimization and sector separation, the resulting $M_\infty$ values are stable within the present framework.
Tables~\ref{tab:charmonium-comparison} and \ref{tab:bottomonium-comparison} list the calibration states. 
Residual statistics $\delta M_i=M_i^{\rm th}-M_i^{\rm exp}$ give an ${\rm RMS}$ $=34$~($24$)~$\mathrm{MeV}$ and the standard error $\sigma_{\delta M}=34$~($23$)~$\mathrm{MeV}$ for $c\bar{c}$~($b\bar{b}$), improving to ${\rm RMS}\simeq10$~($4$)~$\mathrm{MeV}$ for low-lying states, which we take as the uncertainty on predicted masses and on $M_\infty$.

\renewcommand{\tabcolsep}{0.20cm}
\renewcommand{\arraystretch}{1.1}
\begin{table}[!htbp]
\caption{Final fitting results of the parameters in the screened GI model. The last column lists the parameters from the unified heavy-meson fit, 
    while the charmonium and bottomonium columns give the final production values after the system-dependent refits.}
	\label{tab:params}
	\begin{tabular*}{\linewidth}{@{\extracolsep{\fill}}lccc}
		\toprule[1.0pt]\toprule[1.0pt]
		Parameters & Charmonium & Bottomonium & Heavy meson \\
		\midrule[0.75pt]
        $m_n\ ({\rm GeV})$          & \multicolumn{2}{c}{\dots} & 0.456 \\
        $m_s\ ({\rm GeV})$          & \multicolumn{2}{c}{\dots} & 0.617 \\
        $m_c\ ({\rm GeV})$          & \multicolumn{2}{c}{1.805} & 1.805 \\
        $m_b\ ({\rm GeV})$          & \multicolumn{2}{c}{5.151} & 5.151 \\
        $c\ ({\rm GeV})$            & \multicolumn{2}{c}{-0.648} & -0.648 \\
        $\sigma_0\ ({\rm GeV})$     & \multicolumn{2}{c}{1.771} & 1.771 \\
        $s$                         & \multicolumn{2}{c}{1.146} & 1.146 \\
        $b\ ({\rm GeV}^2)$          & 0.256 & 0.247 & 0.252 \\
        $\mu\ ({\rm GeV})$          & 0.144 & 0.122 & 0.135 \\
        $\epsilon_{\rm Coul}$       & 0.0 & 0.0 & 0.0 \\
        $\epsilon_{\rm cont}$       & -0.359 & -0.498 & -0.320 \\
        $\epsilon_{\rm so(\nu)}$   & -0.499 & -0.146 & -0.343 \\
        $\epsilon_{\rm so(s)}$      & 1.000 & -0.502 & 1.000 \\
        $\epsilon_{\rm tens}$       & -0.500 & -0.846 & -0.500 \\
		\bottomrule[1.0pt]\bottomrule[1.0pt]
	\end{tabular*}
\end{table}

\renewcommand{\tabcolsep}{0.3cm}
\renewcommand{\arraystretch}{1.1}
\begin{table}[!htbp]
    \caption{Charmonium calibration spectrum.
    Masses are in MeV. $M_{\rm err}$ is the experimental uncertainty (rounded up to the nearest MeV).}
    \label{tab:charmonium-comparison}
    \begin{tabular*}{\linewidth}{@{\extracolsep{\fill}}cccccl}
        \toprule[1.0pt]
        \toprule[1.0pt]
        $n^{2S+1}L_J$ & $J^{PC}$ & $M_{\rm th}$ & $M_{\rm exp}$ & $M_{\rm err}$ & State \\
        \midrule[0.75pt]
        $1^{1}S_{0}$ & $0^{-+}$ & $2990$ & $2984$ & $1$ & $\eta_c(1S)$ \\
        $2^{1}S_{0}$ & $0^{-+}$ & $3627$ & $3638$ & $1$ & $\eta_c(2S)$ \\
        $1^{3}S_{1}$ & $1^{--}$ & $3103$ & $3097$ & $1$ & $J/\psi(1S)$ \\
        $2^{3}S_{1}$ & $1^{--}$ & $3672$ & $3686$ & $1$ & $\psi(2S)$ \\
        $3^{3}S_{1}$ & $1^{--}$ & $4015$ & $4040$ & $4$ & $\psi(4040)$ \\
        $4^{3}S_{1}$ & $1^{--}$ & $4252$ & $4222$ & $3$ & $\psi(4230)$ \\
        $5^{3}S_{1}$ & $1^{--}$ & $4422$ & $4415$ & $5$ & $\psi(4415)$ \\
        $1^{1}P_{1}$ & $1^{+-}$ & $3522$ & $3525$ & $1$ & $h_c(1P)$ \\
        $1^{3}P_{0}$ & $0^{++}$ & $3419$ & $3415$ & $1$ & $\chi_{c0}(1P)$ \\
        $1^{3}P_{1}$ & $1^{++}$ & $3506$ & $3511$ & $1$ & $\chi_{c1}(1P)$ \\
        $1^{3}P_{2}$ & $2^{++}$ & $3563$ & $3556$ & $1$ & $\chi_{c2}(1P)$ \\
        $2^{3}P_{2}$ & $2^{++}$ & $3935$ & $3923$ & $1$ & $\chi_{c2}(3930)$ \\
        $1^{3}D_{1}$ & $1^{--}$ & $3787$ & $3774$ & $1$ & $\psi(3770)$ \\
        $2^{3}D_{1}$ & $1^{--}$ & $4083$ & $4191$ & $5$ & $\psi(4160)$ \\
        $3^{3}D_{1}$ & $1^{--}$ & $4296$ & $4374$ & $7$ & $\psi(4360)$ \\
        $1^{3}D_{2}$ & $2^{--}$ & $3819$ & $3824$ & $1$ & $\psi_2(3823)$ \\
        $1^{3}D_{3}$ & $3^{--}$ & $3843$ & $3843$ & $1$ & $\psi_3(3842)$ \\
        \bottomrule[1.0pt]
        \bottomrule[1.0pt]
    \end{tabular*}
\end{table}

\renewcommand{\tabcolsep}{0.3cm}
\renewcommand{\arraystretch}{1.1}
\begin{table}[!htbp]
    \caption{Bottomonium calibration spectrum.
    Masses are in MeV. $M_{\rm err}$ is the experimental uncertainty (rounded up to the nearest MeV).}
    \label{tab:bottomonium-comparison}
    \begin{tabular*}{\linewidth}{@{\extracolsep{\fill}}cccccl}
        \toprule[1.0pt]
        \toprule[1.0pt]
        $n^{2S+1}L_J$ & $J^{PC}$ & $M_{\rm th}$ & $M_{\rm exp}$ & $M_{\rm err}$ & State \\
        \midrule[0.75pt]
        $1^{1}S_{0}$ & $0^{-+}$ & $9400$ & $9399$ & $2$ & $\eta_b(1S)$ \\
        $2^{1}S_{0}$ & $0^{-+}$ & $9996$ & $9999$ & $4$ & $\eta_b(2S)$ \\
        $1^{3}S_{1}$ & $1^{--}$ & $9458$ & $9460$ & $1$ & $\Upsilon(1S)$ \\
        $2^{3}S_{1}$ & $1^{--}$ & $10021$ & $10023$ & $1$ & $\Upsilon(2S)$ \\
        $3^{3}S_{1}$ & $1^{--}$ & $10358$ & $10355$ & $1$ & $\Upsilon(3S)$ \\
        $4^{3}S_{1}$ & $1^{--}$ & $10604$ & $10579$ & $2$ & $\Upsilon(4S)$ \\
        $5^{3}S_{1}$ & $1^{--}$ & $10797$ & $10885$ & $3$ & $\Upsilon(10860)$ \\
        $6^{3}S_{1}$ & $1^{--}$ & $10958$ & $11000$ & $4$ & $\Upsilon(11020)$ \\
        $1^{1}P_{1}$ & $1^{+-}$ & $9892$ & $9899$ & $1$ & $h_b(1P)$ \\
        $2^{1}P_{1}$ & $1^{+-}$ & $10259$ & $10260$ & $2$ & $h_b(2P)$ \\
        $1^{3}P_{0}$ & $0^{++}$ & $9856$ & $9859$ & $1$ & $\chi_{b0}(1P)$ \\
        $2^{3}P_{0}$ & $0^{++}$ & $10236$ & $10233$ & $1$ & $\chi_{b0}(2P)$ \\
        $1^{3}P_{1}$ & $1^{++}$ & $9886$ & $9893$ & $1$ & $\chi_{b1}(1P)$ \\
        $2^{3}P_{1}$ & $1^{++}$ & $10256$ & $10256$ & $1$ & $\chi_{b1}(2P)$ \\
        $3^{3}P_{1}$ & $1^{++}$ & $10522$ & $10513$ & $1$ & $\chi_{b1}(3P)$ \\
        $1^{3}P_{2}$ & $2^{++}$ & $9907$ & $9912$ & $1$ & $\chi_{b2}(1P)$ \\
        $2^{3}P_{2}$ & $2^{++}$ & $10270$ & $10269$ & $1$ & $\chi_{b2}(2P)$ \\
        $3^{3}P_{2}$ & $2^{++}$ & $10532$ & $10524$ & $1$ & $\chi_{b2}(3P)$ \\
        $1^{1}D_{2}$ & $2^{-+}$ & $10165$ & $10164$ & $2$ & $\Upsilon_2(1D)$ \\
        \bottomrule[1.0pt]
        \bottomrule[1.0pt]        
    \end{tabular*}
\end{table}

\begin{figure*}[!htbp]
    \centering
    \subfigure[\(\ \)Energy levels]{\includegraphics[width=0.47\textwidth]{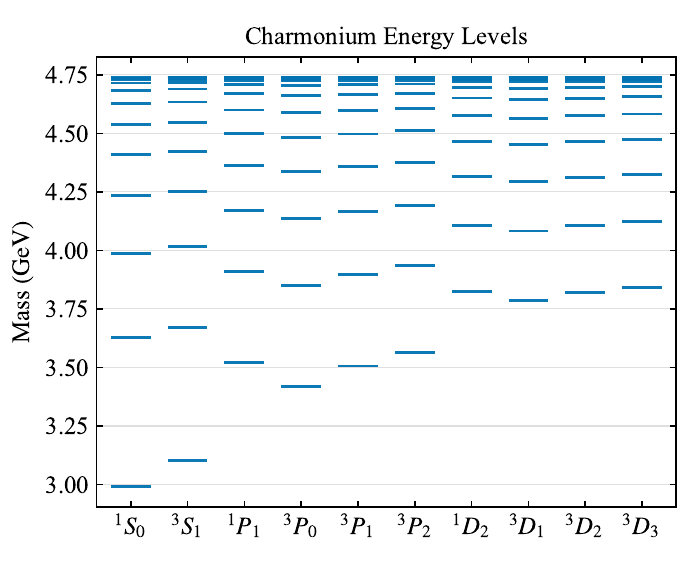}}
    \subfigure[\(\ \)Mass and radius bands]{\includegraphics[width=0.47\textwidth]{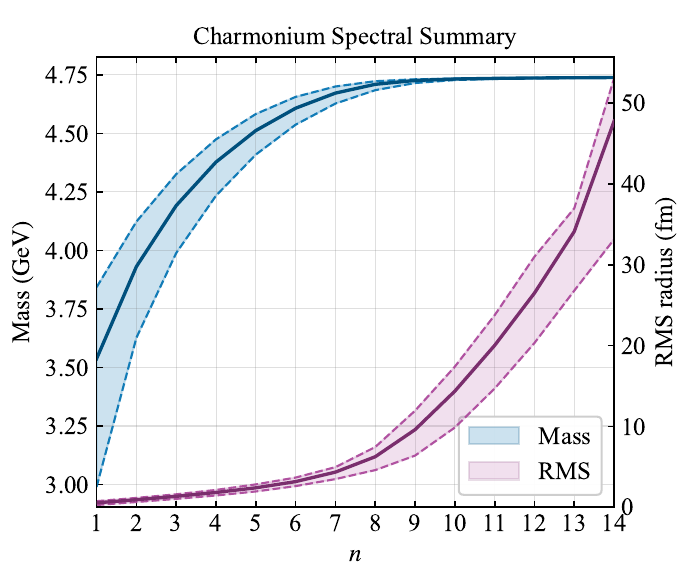}}
    \caption{The calculated masses and mass--radius bands in the charmonium system. The energy levels for the $nS$, $nP$, and $nD$ families are presented in subfigure (a). In subfigure (b), the upper, lower, and average values across channels are presented. The mass band narrows while the RMS-radius band widens with increasing radial quantum number $n$, demonstrating the mass--radius decoupling that defines the operational upper limit.}
    \label{fig:charmonium}
\end{figure*}

Masses and RMS radii for all $nS$, $nP$, and $nD$ channels up to $n=14$ are then computed with the production parameters.
Two global features stand out as generic properties of highly radially excited solutions for masses and RMS radii.
First, in every channel the masses approach a plateau fixed by the asymptotic screened potential,
\begin{equation}
    M_\infty = m_1 + m_2 + \frac{b}{\mu} + c,
    \label{equ:Minf}
\end{equation}
which lies at $4.74\,\mathrm{GeV}$ for charmonium and $11.67\,\mathrm{GeV}$ for bottomonium.
Second, the RMS radii continue to increase without bound, reaching $30$--$50\,\mathrm{fm}$ (charmonium) and $12$--$15\,\mathrm{fm}$ (bottomonium) at $n=14$.
Figure~\ref{fig:charmonium} displays the calculated charmonium energy levels and the corresponding channel-averaged masses and radii.
The spectrum therefore exhibits clear mass truncation accompanied by spatial divergence---the mass--radius decoupling in which additional radial nodes barely shift the mass while the spatial size keeps growing. 

To quantify this decoupling, we monitor the neighboring mass gaps and radius increments, defined by the successive differences
\begin{equation}
    \Delta M_n=M_n-M_{n-1}, \quad \Delta R_n=R_n-R_{n-1}.
\end{equation}
The operational upper limit is identified by the point at which neighboring mass gaps fall below \(10\,\mathrm{MeV}\) while RMS radii still increase by several fm. 
Explicitly, the caution index \(n_{\rm caution}\) is the smallest \(n\) satisfying the thresholds:
\begin{equation}
    \begin{aligned}
        &\min(\Delta M_n,\Delta M_{n+1}) < 10\,\mathrm{MeV}, \\
        &\max(\Delta R_n,\Delta R_{n+1}) > 3\,\mathrm{fm} \quad (c\bar{c}), \\
        &\max(\Delta R_n,\Delta R_{n+1}) > 2\,\mathrm{fm} \quad (b\bar{b}), 
    \end{aligned}
    \label{equ:threshold-based}
\end{equation}
with the safe range ending at \(n_{\rm safe}=n_{\rm caution}-1\). The resulting channel-by-channel boundaries are listed in Table~\ref{tab:limit}. 
Across all channels, charmonium reaches the truncation region earlier than bottomonium. The $D$-wave states are the first to enter the caution regime ($n=8$), 
followed by the $P$-wave ($n=9$) and $S$-wave ($n=10$) families. In bottomonium the corresponding transitions occur at $n=12$ ($D$) and $n=13$ ($S,P$). 
States beyond $n_{\rm safe}$ therefore cannot be regarded as observable quarkonia within the present analysis.

\renewcommand{\tabcolsep}{0.20cm}
\renewcommand{\arraystretch}{1.1}
\begin{table}[!htbp]
    \caption{Operational upper limits for charmonium and bottomonium.
    $n_{\rm safe}$ is the last reliable radial quantum number;
    at $n_{\rm caution}$ the mass gaps have collapsed below $10\,\mathrm{MeV}$
    while the radii are still growing rapidly.
    $\Delta M$ (MeV) and $\Delta R$ (fm) are the neighboring mass gap and RMS-radius
    increment at the caution onset. $R_{\rm caution}$ (fm) is the RMS radius of the caution state.}
    \label{tab:limit}
    \begin{tabular*}{\linewidth}{@{\extracolsep{\fill}}llccccc}
        \toprule[1.0pt]
        \toprule[1.0pt]
        System & Channel & $n_{\rm safe}$ & $n_{\rm caution}$ & $\Delta M$ & $\Delta R$ & $R_{\rm caution}$ \\
        \midrule[0.75pt]
        \multirow{10}{*}{$c\bar{c}$}
            & $^1S_0$ & 9 & 10 & 4.8 & 4.8 & 9.9 \\
            & $^3S_1$ & 9 & 10 & 4.2 & 4.9 & 10.7 \\
            & $^1P_1$ & 8 & 9  & 5.7 & 4.7 & 9.2 \\
            & $^3P_0$ & 8 & 9  & 6.4 & 4.6 & 8.7 \\
            & $^3P_1$ & 8 & 9  & 5.8 & 4.7 & 9.1 \\
            & $^3P_2$ & 8 & 9  & 5.5 & 4.8 & 9.3 \\
            & $^1D_2$ & 7 & 8  & 8.9 & 4.4 & 7.3 \\
            & $^3D_1$ & 7 & 8  & 9.9 & 4.2 & 7.0 \\
            & $^3D_2$ & 7 & 8  & 9.1 & 4.4 & 7.2 \\
            & $^3D_3$ & 7 & 8  & 8.2 & 4.5 & 7.5 \\
        \midrule[0.75pt]
        \multirow{10}{*}{$b\bar{b}$}
            & $^1S_0$ & 12 & 13 & 9.0 & 3.2 & 10.4 \\
            & $^3S_1$ & 12 & 13 & 9.0 & 3.7 & 12.0 \\
            & $^1P_1$ & 12 & 13 & 6.9 & 2.1 & 9.7 \\
            & $^3P_0$ & 12 & 13 & 6.9 & 2.4 & 10.2 \\
            & $^3P_1$ & 12 & 13 & 6.9 & 2.1 & 9.7 \\
            & $^3P_2$ & 12 & 13 & 6.9 & 2.8 & 10.8 \\
            & $^1D_2$ & 11 & 12 & 5.5 & 2.3 & 9.0 \\
            & $^3D_1$ & 12 & 13 & 2.6 & 2.2 & 9.8 \\
            & $^3D_2$ & 11 & 12 & 5.5 & 2.3 & 8.9 \\
            & $^3D_3$ & 11 & 12 & 5.5 & 2.6 & 9.8 \\
        \bottomrule[1.0pt]\bottomrule[1.0pt]
    \end{tabular*}
\end{table}

Two independent diagnostics corroborate the threshold-based boundary. A monotonic indicator is defined by
\begin{equation}
    f_n = \log_{10}\Bigl[\frac{\Delta R_n + \epsilon_r}{\Delta M_n + \epsilon_M}\Bigr].
\end{equation}
To locate the transition, two changepoint procedures are applied to the sequence \(\{f_n\}\). 
The cumulative-sum method follows the running sum of standardized deviations from the mean, 
while the two-segment method identifies the division that minimizes the total residual sum of squares in a piecewise-constant fit. 
Both place the structural transition at \(n\sim 8\)--\(10\) for charmonium and \(n\sim 9\)--\(11\) for bottomonium, earlier in the lighter system. 
The dimensionless resolvability is given by
\begin{equation}
    \chi_n = \Delta M_n R_n,
\end{equation}
with \(\Delta M\) in GeV and \(R\) in GeV\(^{-1}\). It remains roughly constant and then falls rapidly below unity at or above \(n_{\rm caution}\) in every channel, 
showing that neighboring levels are no longer physically resolvable. All three diagnostics converge on the same truncation picture. 
Full analysis and comparisons appear in the Supplemental Material.

Beyond the mass--radius decoupling picture, the absolute spatial extent at the caution boundary provides a direct physical criterion. 
The channel-by-channel results in Table~\ref{tab:limit} show that the RMS radii at the caution onset lie in the range $7$--$11\,\mathrm{fm}$ for charmonium and $9$--$12\,\mathrm{fm}$ for bottomonium, clustering around $10\,\mathrm{fm}$. 
This is an order of magnitude larger than the characteristic QCD confinement scale of \(\sim1\,\mathrm{fm}\). 
At scales far exceeding the string-breaking distance, the linear potential has already been saturated by light-quark pair creation,
where the spectra and internal hadron structure are constrained by color confinement.
Consequently, the caution boundary fixed by the thresholds in Eq.~(\ref{equ:threshold-based}) marks the onset of mass--radius decoupling, consistent with this spatial criterion.

In this work, we have performed the first quantitative determination of the
operational upper radial limits of heavy quarkonia by solving a screened,
relativized GI Hamiltonian using the GEM.
A multi-staged $\chi^2$ fit fixes the production parameters and yields
limiting masses of \(4.74\,\mathrm{GeV}\) (\(c\bar{c}\)) and
\(11.67\,\mathrm{GeV}\) (\(b\bar{b}\)), while the root-mean-square radii
grow to values an order of magnitude above the confinement scale. A
threshold-based mass--gap criterion, corroborated with changepoint analysis and
resolvability collapse, places the limiting radial quantum numbers at
\(n\approx 8\)--\(10\) for \(c\bar{c}\) and \(n\approx 12\)--\(13\) for
\(b\bar{b}\). Beyond these boundaries, the conventional \(q\bar{q}\)
bound-state description ceases to be valid.

These findings address the long-standing question of whether the radial
excitation ladder of quarkonia truncates, providing a concrete physical
boundary anchored to the QCD confinement scale. The mass--radius decoupling
criterion is robust and is directly testable with forthcoming
high-statistics data from BESIII, Belle II, and LHCb in the regions above
\(4.7\,\mathrm{GeV}\) and \(11\,\mathrm{GeV}\).

\begin{acknowledgments}
This work is supported by the Natural Science Foundation of Gansu Province (No.~26RCKA012, No.~25JRRA799), 
the National Natural Science Foundation of China under Grants No.~12335001 and No.~12247101, the ``111 Center'' under Grant No.~B20063, 
the Fundamental Research Funds for the Central Universities (lzujbky-2023-stlt01), and Lanzhou City High-Level Talent Funding.

AI Usage Declaration: The initial version of the manuscript was drafted with the assistance of Grok Build 0.1 \cite{xaigrok}.
The Supplemental Material was initially prepared by GPT-5.4 \cite{openaigpt} and subsequently revised by the authors.
Data processing, visualization, and the truncation analysis were carried out with GPT-5.4.
The authors reviewed and edited the output as needed and take full responsibility for the content of the published article.

Data Availability: The codebase and numerical tests are available at GitHub \cite{gemstore}.
\end{acknowledgments}

%

\clearpage
\onecolumngrid
\setcounter{section}{0}
\setcounter{subsection}{0}
\setcounter{subsubsection}{0}
\setcounter{equation}{0}
\setcounter{figure}{0}
\setcounter{table}{0}
\setcounter{secnumdepth}{3}
\renewcommand{\thesection}{S\arabic{section}}
\renewcommand{\thesubsection}{S\arabic{section}.\arabic{subsection}}
\renewcommand{\theequation}{S\arabic{equation}}
\renewcommand{\thefigure}{S\arabic{figure}}
\renewcommand{\thetable}{S\arabic{table}}
\begin{center}
{\large\bfseries Supplemental Material}
\end{center}
\vspace{1em}
\input{support_body.tex}

\end{document}

%% file: support_body.tex
This Supplemental Material collects the technical derivations, numerical procedures, and extended results underlying the main manuscript. 
It includes the screen-modified Godfrey--Isgur model, the Gaussian expansion method, matrix-element evaluation and stabilization details, 
the multi-stage parameter fitting procedure, the extended charmonium and bottomonium spectra, and the limitation analysis for highly excited states. 
Sections~\ref{sec:method} and \ref{sec:fit} present the theoretical framework and fitting strategy, 
Sec.~\ref{sec:results} summarizes the extended spectra together with the limitation analysis and additional diagnostics, 
and the final section gives a brief summary of the supplemental results.

\section{Theoretical Framework}
\label{sec:method}

This section presents the theoretical and numerical ingredients underlying the spectroscopy computation. 
We first specify the screened Godfrey--Isgur Hamiltonian, then introduce the Gaussian expansion basis used to represent the quarkonium wave functions, 
and finally describe the orthogonal-basis construction and Gaussian quadrature algorithms employed in the evaluation of the matrix elements.

\subsection{Screen-modified Godfrey--Isgur Hamiltonian}
\label{sec:model}

In the main text we use a screen-modified Godfrey--Isgur quark model to describe heavy quarkonium for a wide range of radial excitations. 
The purpose of this subsection is to state the Hamiltonian in full and clarify how screening, smearing, and relativistic corrections are implemented in the numerical calculations. 
For compact $c\bar c$ and $b\bar b$ states, the original GI model with linear confinement already provides an accurate description of the short-distance dynamics. 
The present work is concerned with much larger radial excitations, for which the long-distance interaction must be saturated to mimic the onset of string breaking and unquenched effects.

The Hamiltonian of the MGI model can be written as
\begin{equation}
    H = \sum_{i} \sqrt{p_i^2+m_i^2} + \sum_{i<j} (V_{ij}^{\textrm{Coul}} + V_{ij}^{\textrm{conf}} + V_{ij}^{\textrm{cont}}
        + V_{ij}^{\textrm{so}(\nu)} + V_{ij}^{\textrm{so}(s)} + V_{ij}^{\textrm{tens}}), \label{equ:HamiltonianGI}
\end{equation}
where $\sqrt{p_i^2+m_i^2}$ is the relativistic kinetic energy of quark $i$, and $V_{ij}^{\textrm{Coul}}$, $V_{ij}^{\textrm{conf}}$,
$V_{ij}^{\textrm{cont}}$, $V_{ij}^{\textrm{so}(\nu)}$, $V_{ij}^{\textrm{so}(s)}$, and $V_{ij}^{\textrm{tens}}$ are the color-Coulomb,
confinement, contact, spin-orbit, Thomas precession, and tensor interactions between quarks $i$ and $j$, respectively.
This Hamiltonian is a direct generalization of the original GI model, but with the confining term modified to a screened form. 
The remaining terms are taken from the GI model directly.

The central modification relative to a purely static potential is the use of smeared interactions. 
In the relativized GI framework, the interaction is not evaluated as if the quarks were pointlike sources at a fixed separation. 
Instead, relativistic motion is incorporated by folding the bare potential with a Gaussian distribution, which represents the finite spread of the quark coordinates. 
This procedure softens the short-distance singularity of the Coulomb term and yields a more realistic effective interaction, 
while preserving the overall structure of the potential model. For the color-Coulomb interaction, the smeared form $\widetilde{G}(r)$ is therefore written as
\begin{align}
    \widetilde{G}(r) &= \int d^3\mathbf{r^{\prime}} \big(\frac{\sigma^{3}}{\pi^{3/2}}
    e^{-\sigma^2 (\mathbf{r}-\mathbf{r^{\prime}})^2}\big) \big(-\frac{4}{3}\frac{\alpha}{r^{\prime}}\big) \nonumber \\
    &= \iint d\theta dr^{\prime} 2\pi{r^{\prime}}^2\textrm{sin}(\theta) \big(\frac{\sigma^{3}}{\pi^{3/2}}
    e^{-\sigma^2 (r^2+{r^{\prime}}^2-2rr^{\prime}\textrm{cos}(\theta))}\big) \big(-\frac{4}{3}\frac{\alpha}{r^{\prime}}\big) \nonumber \\
    &= -\frac{4}{3}\frac{\alpha}{r} \textrm{erf}(\sigma r), \label{equ:smearingGr}
\end{align}
while the corresponding smeared confinement interaction $\widetilde{S}(r)$ is given by
\begin{align}
    \widetilde{S}(r) &= \int d^3\mathbf{r^{\prime}} \big(\frac{\sigma^{3}}{\pi^{3/2}}
    e^{-\sigma^2 (\mathbf{r}-\mathbf{r^{\prime}})^2}\big) \big(\frac{b(1-e^{-\mu r^{\prime}})}{\mu}+c\big) \nonumber \\
    &= \iint d\theta dr^{\prime} 2\pi{r^{\prime}}^2\textrm{sin}(\theta) \big(\frac{\sigma^{3}}{\pi^{3/2}}
    e^{-\sigma^2 (r^2+{r^{\prime}}^2-2rr^{\prime}\textrm{cos}(\theta))}\big) \big(\frac{b(1-e^{-\mu r^{\prime}})}{\mu}+c\big) \nonumber \\
    &= \frac{be^{-\mu r}}{4r\mu\sigma^2}\Big[4r\sigma^2 e^{\mu r} + (\mu-2r\sigma^2)e^{\mu^2/(4\sigma^2)} \textrm{erfc}\big(\frac{\mu-2r\sigma^2}{2\sigma}\big)
    - (\mu+2r\sigma^2)e^{\mu^2/(4\sigma^2)+2\mu r} \textrm{erfc}\big(\frac{\mu+2r\sigma^2}{2\sigma}\big)\Big] + c. \label{equ:smearingSr}
\end{align}
Here, $\sigma$ is the smearing parameter depending on the quark masses $m_i$ and $m_j$, while $\mu$ denotes the screening parameter for the confinement interaction.
\begin{figure}[!htbp]
    \centering
    \subfigure[Smearing of the color-Coulomb interaction.]{
        \includegraphics[width=0.47\textwidth]{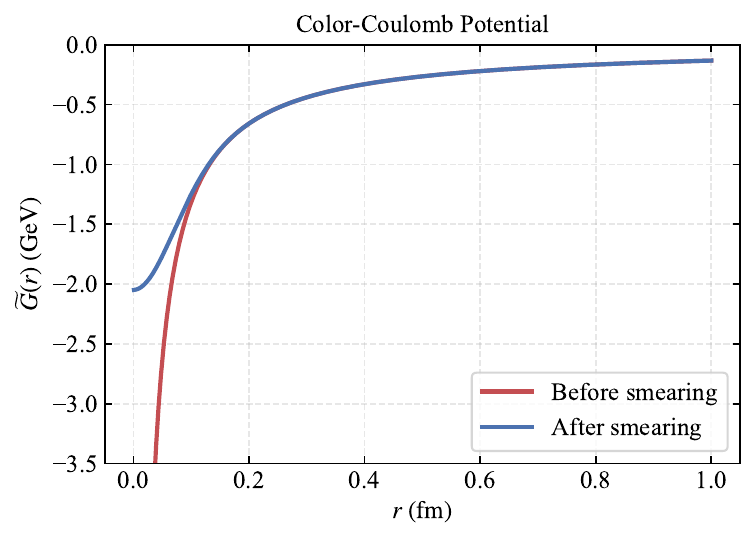}
    }
    \subfigure[Smearing of the screened confinement interaction.]{
        \includegraphics[width=0.47\textwidth]{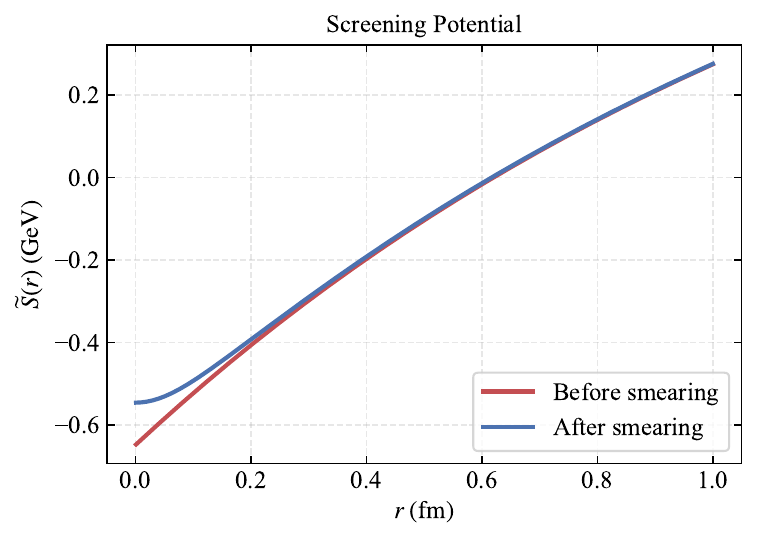}
    }
    \caption{Illustration of the smearing procedure for $\widetilde{G}(r)$ and $\widetilde{S}(r)$. 
    The Gaussian convolution removes the short-range singularity of the color-Coulomb potential and smooths the screened confinement potential, 
    while preserving the overall structure of the potential model.}
    \label{fig:smearing-potentials}
\end{figure}
Fig.~\ref{fig:smearing-potentials} demonstrates the physical effect of the Gaussian convolution on the two central components of the potential, 
showing the short-range smearing behavior while the overall interaction pattern is maintained.
For numerical purposes, Eqs. (\ref{equ:smearingGr}) and (\ref{equ:smearingSr}) can be written in terms of $r_{ij}$:
\begin{equation}
    \widetilde{G}_{ij} = C_{ij}\frac{\alpha}{r_{ij}} \textrm{erf}(\sigma_{ij}r_{ij}), \label{equ:smearedGij}
\end{equation}
\begin{equation}
    \widetilde{S}_{ij} = -\frac{3}{4}C_{ij}\Bigg\{\frac{be^{-\mu r_{ij}}}{4r_{ij}\mu\sigma_{ij}^2}\Bigg[4r_{ij}\sigma_{ij}^2 e^{\mu r_{ij}} + (\mu-2r_{ij}\sigma_{ij}^2)e^{\mu^2/(4\sigma_{ij}^2)} \textrm{erfc}\big(\frac{\mu-2r_{ij}\sigma_{ij}^2}{2\sigma_{ij}}\big) \\
    - (\mu+2r_{ij}\sigma_{ij}^2)e^{\mu^2/(4\sigma_{ij}^2)+2\mu r_{ij}} \textrm{erfc}\big(\frac{\mu+2r_{ij}\sigma_{ij}^2}{2\sigma_{ij}}\big)\Bigg] + c\Bigg\}, \label{equ:smearedSij}
\end{equation}
Here, $C_{ij}$ denotes the color factor for mesons as $C_{ij} = -4/3$, and the smearing width $\sigma_{ij}$ depends on the quark masses through
\begin{equation}
    \sigma_{ij}^2 = \sigma_0^2 \big[\frac{1}{2} + \frac{1}{2}(\frac{4m_i m_j}{(m_i+m_j)^2})^4\big] + s^2 \big(\frac{2m_i m_j}{m_i+m_j}\big)^2, \label{equ:sigmaij}
\end{equation}
with free parameters $\sigma_0$ and $s$. 
Finally, to reproduce the running behavior of the strong coupling constant for asymptotic freedom,
we write the Coulomb term (\ref{equ:smearedGij}) in combination with series of parameters $\alpha_k$ and $\gamma_k$:
\begin{equation}
    \widetilde{G}_{ij} = C_{ij} \sum_{k} \frac{\alpha_{k}}{r_{ij}} \textrm{erf}(\sigma_{kij}r_{ij}), \label{equ:runningSmearedGij}
\end{equation}
where $\sigma_{kij}$ is defined by the smearing parameter $\sigma_{ij}$ as $\sigma_{kij}^{-2} = \sigma_{ij}^{-2} + \gamma_k^{-2}$.
The parameters $\alpha_k$ and $\gamma_k$ have been determined to be $\alpha_k=\{0.25,\ 0.15,\ 0.20\}$ and $\gamma_k=\{1/2,\ \sqrt{5/2},\ 5\sqrt{10}\}$, respectively, 
by fitting the running coupling $\alpha_s(Q^2) = \sum_{k} \alpha_{k} e^{-Q^2/4\gamma_{k}}$ to the lowest-order QCD with an effective scale parameter $\Lambda_{\textrm{QCD}}=200\ $MeV.

Beyond the smearing of the central Coulomb and confinement potentials, the GI framework also incorporates momentum-dependent factors in the spin-dependent sector. 
These factors account for the fact that spin-orbit, contact, and tensor interactions are modified by the finite quark momentum, and therefore cannot be described by purely static operators. 
In this way, the central smearing and the momentum-dependent corrections together provide a more realistic effective interaction.
The corresponding factors are denoted by $\beta_{ij}$, $\delta_{ij}$, $\delta_{ii}$, and $\delta_{jj}$, and defined as
\begin{align}
    \beta_{ij} &= 1 + \frac{p_{ij}^2}{(p_{ij}^2+m_{i}^2)^{1/2} (p_{ij}^2+m_{j}^2)^{1/2}}, \label{equ:betaij} \\
    \delta_{ij} &= \frac{m_{i} m_{j}}{(p_{ij}^2+m_{i}^2)^{1/2} (p_{ij}^2+m_{j}^2)^{1/2}}, \label{equ:deltaij} \\
    \delta_{ii} &= \frac{m_{i} m_{i}}{(p_{ii}^2+m_{i}^2)^{1/2} (p_{ii}^2+m_{i}^2)^{1/2}}, \label{equ:deltaii} \\
    \delta_{jj} &= \frac{m_{j} m_{j}}{(p_{jj}^2+m_{j}^2)^{1/2} (p_{jj}^2+m_{j}^2)^{1/2}}, \label{equ:deltajj}
\end{align}
where $p_{ij}$ is the momentum of the $ij$ pair in the center-of-mass frame. 
These factors reduce to unity in the nonrelativistic limit and therefore provide a controlled interpolation between nonrelativistic intuition and the relativized GI construction. 
The explicit interaction terms entering Eq.~(\ref{equ:HamiltonianGI}) are given below.

The color-Coulomb potential $V_{ij}^{\textrm{Coul}}$ is given as
\begin{equation}
    V_{ij}^{\textrm{Coul}} = \beta_{ij}^{\frac{1}{2}+\epsilon_{\textrm{Coul}}+\epsilon_{\textrm{Coul}}^2} \widetilde{G}_{ij} \beta_{ij}^{\frac{1}{2}+\epsilon_{\textrm{Coul}}+\epsilon_{\textrm{Coul}}^2}, \label{equ:VijCoul}
\end{equation}
with a parameter $\epsilon_{\textrm{Coul}}$ controlling the strength of the relativistic correction. 
The confining interaction is taken directly from the smeared potential $\widetilde{S}_{ij}$ as Eq.~(\ref{equ:smearedSij}).
The contact contribution, which governs hyperfine splittings in the S-wave sector, can be expressed as
\begin{equation}
    \begin{aligned}
        V_{ij}^{\textrm{cont}} &= \delta_{ij}^{\frac{1}{2}+\epsilon_{\textrm{cont}}+\epsilon_{\textrm{cont}}^2} \frac{2\mathbf{S}_i\cdot\mathbf{S}_j}{3m_i m_j}
        \nabla^2 \widetilde{G}_{ij} \delta_{ij}^{\frac{1}{2}+\epsilon_{\textrm{cont}}+\epsilon_{\textrm{cont}}^2}
        = \delta_{ij}^{\frac{1}{2}+\epsilon_{\textrm{cont}}+\epsilon_{\textrm{cont}}^2} \frac{32e^{-\sigma_{kij}^2 r_{ij}^2}\alpha_{k}\sigma_{kij}^3}{9\sqrt{\pi}m_i m_j} \mathbf{S}_i\cdot\mathbf{S}_j
        \delta_{ij}^{\frac{1}{2}+\epsilon_{\textrm{cont}}+\epsilon_{\textrm{cont}}^2}, \label{equ:Vijcont}
    \end{aligned}
\end{equation}
which follows from the action of the spherical Laplacian operator $\nabla^2=\frac{1}{r^2}\frac{\partial}{\partial r} \left(r^2 \frac{\partial}{\partial r}\right)$ on $\widetilde{G}_{ij}$. 
The remaining fine-structure splittings are controlled by the spin-orbit and tensor operators. The spin-orbit contribution is
\begin{equation}
    \begin{aligned}
        V_{ij}^{\textrm{so}(\nu)} &= \frac{1}{r_{ij}} \frac{\textrm{d}\widetilde{G}_{ij}}{\textrm{d}r_{ij}}
        \big(\delta_{ii}^{\frac{1}{2}+\epsilon_{\textrm{so}(\nu)}+\epsilon_{\textrm{so}(\nu)}^2} \frac{\mathbf{L}\cdot\mathbf{S}_i}{2m_i^2} \delta_{ii}^{\frac{1}{2}+\epsilon_{\textrm{so}(\nu)}+\epsilon_{\textrm{so}(\nu)}^2}
        + \delta_{jj}^{\frac{1}{2}+\epsilon_{\textrm{so}(\nu)}+\epsilon_{\textrm{so}(\nu)}^2} \frac{\mathbf{L}\cdot\mathbf{S}_j}{2m_j^2} \delta_{jj}^{\frac{1}{2}+\epsilon_{\textrm{so}(\nu)}+\epsilon_{\textrm{so}(\nu)}^2}
        + \delta_{ij}^{\frac{1}{2}+\epsilon_{\textrm{so}(\nu)}+\epsilon_{\textrm{so}(\nu)}^2} \frac{\mathbf{L}\cdot\mathbf{S}_i+\mathbf{L}\cdot\mathbf{S}_j}{2m_i m_j} \delta_{ij}^{\frac{1}{2}+\epsilon_{\textrm{so}(\nu)}+\epsilon_{\textrm{so}(\nu)}^2} \big), \label{equ:Vijsonu}
    \end{aligned}
\end{equation}
while the Thomas-precession contribution is
\begin{equation}
    \begin{aligned}
        V_{ij}^{\textrm{so}(s)} &= -\frac{1}{r_{ij}} \frac{\textrm{d}\widetilde{S}_{ij}}{\textrm{d}r_{ij}}
        \big(\delta_{ii}^{\frac{1}{2}+\epsilon_{\textrm{so}(s)}+\epsilon_{\textrm{so}(s)}^2} \frac{\mathbf{L}\cdot\mathbf{S}_i}{2m_i^2} \delta_{ii}^{\frac{1}{2}+\epsilon_{\textrm{so}(s)}+\epsilon_{\textrm{so}(s)}^2}
        + \delta_{jj}^{\frac{1}{2}+\epsilon_{\textrm{so}(s)}+\epsilon_{\textrm{so}(s)}^2} \frac{\mathbf{L}\cdot\mathbf{S}_j}{2m_j^2} \delta_{jj}^{\frac{1}{2}+\epsilon_{\textrm{so}(s)}+\epsilon_{\textrm{so}(s)}^2} \big). \label{equ:Vijsos}
    \end{aligned}
\end{equation}
Finally, the tensor interaction takes the form
\begin{equation}
    \begin{aligned}
        V_{ij}^{\textrm{tens}} &= \delta_{ij}^{\frac{1}{2}+\epsilon_{\textrm{tens}}+\epsilon_{\textrm{tens}}^2} \frac{1}{3m_i m_j}
        \big(\frac{1}{r_{ij}} \frac{\textrm{d}\widetilde{G}_{ij}}{\textrm{d}r_{ij}} - \frac{\textrm{d}^2\widetilde{G}_{ij}}{\textrm{d}r_{ij}^2}\big)
        \big(\frac{3(\mathbf{S}_i\cdot\mathbf{r}_{ij}) (\mathbf{S}_j\cdot\mathbf{r}_{ij})}{r_{ij}^2} - \mathbf{S}_i\cdot\mathbf{S}_j\big)
        \delta_{ij}^{\frac{1}{2}+\epsilon_{\textrm{tens}}+\epsilon_{\textrm{tens}}^2}. \label{equ:Vijtens}
    \end{aligned}
\end{equation}
The quadratic term in the exponent may be regarded as a simple higher-order correction motivated by a Taylor expansion in the relativistic modification parameter, 
so that the next-order contribution is retained.

Taken together, Eqs.~(\ref{equ:VijCoul})--(\ref{equ:Vijtens}) specify the screen-modified GI Hamiltonian used in this work. 
The parameters $b$, $c$, $\sigma_0$, $s$, the constituent quark masses, and the exponents $\epsilon_{\rm Coul}$, $\epsilon_{\rm cont}$, 
$\epsilon_{\rm so(\nu)}$, $\epsilon_{\rm so(s)}$, and $\epsilon_{\rm tens}$ are determined through the fitting procedure described in Sec.~\ref{sec:fit}. 
Once these inputs are fixed, the remaining step is to solve the resulting eigensystem problem within the GEM basis, as discussed in the next subsection.

\subsection{Gaussian Expansion Method}
\label{sec:gem}

With the MGI Hamiltonian specified, the spectrum is obtained by solving the eigenvalue equation $H|\psi\rangle = E|\psi\rangle$ within a variational framework. 
For mesons, this approach is particularly effective because the color and flavor structures are fixed, and the nontrivial dynamics is carried by the spin and orbital degrees of freedom of the relative motion. 
In this context, the Gaussian Expansion Method provides a promising framework. Its simple analytic form and geometrically spaced scales allow the same basis to describe both short-range behavior and long-range multi-node structure.
We therefore express the full meson state as a direct product of color $|\psi^{color}\rangle$, flavor $|\psi^{flavor}\rangle$, 
spin $|\psi_{s}^{spin}\rangle$, and orbital $|\psi_{l}^{orbit}\rangle$ components, and expand the orbital part through the Rayleigh--Ritz method as
\begin{equation}
    \psi_{l}^{orbit}=\sum_{n=1}^{N_{max}} C_n \phi_{nl}.
\end{equation}
The basis functions are taken as
\begin{equation}
    \begin{aligned}
        \phi_{nl}(\mathbf{r}) &= R_{nl}^{r} Y_{lm}(\mathbf{r}) = N_{nl}^{r} r^{l} e^{-\nu_n r^2} Y_{lm}(\mathbf{r}), \\
        \phi_{nl}(\mathbf{p}) &= R_{nl}^{p} Y_{lm}(\mathbf{p}) = N_{nl}^{p} p^{l} e^{-\frac{p^2}{4\nu_n}} Y_{lm}(\mathbf{p}) \label{equ:GEMbasis}
    \end{aligned}
\end{equation}
with the normalization constants
\begin{equation}
    N_{nl}^{r} = \big(\frac{2^{l+2}(2\nu_n)^{l+\frac{3}{2}}}{\sqrt{\pi}(2l+1)!!}\big)^{\frac{1}{2}}, \quad
    N_{nl}^{p} = (-i)^l \big(\frac{2^{l+2}(2\nu_n)^{-l-\frac{3}{2}}}{\sqrt{\pi}(2l+1)!!}\big)^{\frac{1}{2}}, \label{equ:GEMnormalization}
\end{equation}
and mathematical relations
\begin{equation}
    \begin{aligned}
        (2l+1)!! &= (2l+1)(2l-1)(2l-3)\cdots3\cdot1 = \frac{2^{l+1}\Gamma(l+\frac{3}{2})}{\sqrt{\pi}} = \frac{(2l+1)!}{2^{l} l!}. \label{equ:doublefactorial}
    \end{aligned}
\end{equation}
The Gaussian scale parameters $\nu_n$ are chosen in geometric progression,
\begin{equation}
    \nu_n = \frac{1}{r_n^2}, \quad r_n = r_{min} a^{n-1}, \quad a = \big(\frac{r_{max}}{r_{min}}\big)^{\frac{1}{N_{max}-1}}, \label{equ:GEMsizeparameter}
\end{equation}
where $r_{min}$ and $r_{max}$ denote the smallest and largest length scales for the basis, and $N_{max}$ is the total number of basis functions. 
This geometric construction is more efficient than a linearly spaced grid, as it distributes the basis functions over multiple length scales in a balanced way. 
As a result, the radial region is covered more uniformly, allowing both short-range behavior and long-range structure to be described within the same basis set.

Since the Gaussian basis functions in Eq.~(\ref{equ:GEMbasis}) are not mutually orthogonal, the variational method takes the form of a generalized eigenvalue problem instead of an ordinary matrix diagonalization:
\begin{equation}
    \sum_{n=1}^{N_{max}} \big(\langle \phi_{m}|H|\phi_{n}\rangle - E \langle \phi_{m}|\phi_{n}\rangle\big) C_n = 0, \label{equ:generalizedeigenvalue}
\end{equation}
whose solutions determine the eigenvalues $E$ and the eigenvector components $C_n$. The Hamiltonian and overlap matrices are therefore defined by
\begin{equation}
    H_{mn} = \langle \phi_{m}|H|\phi_{n}\rangle = \int r^2 dr R_{ml}^{r} H R_{nl}^{r}, \quad
    N_{mn} = \langle \phi_{m}|\phi_{n}\rangle = \int r^2 dr R_{ml}^{r} R_{nl}^{r}, \label{equ:GEMmatrixelement}
\end{equation}
where the orthogonality relation of spherical harmonics $Y_{lm}(\mathbf{r})$ has been applied as
\begin{equation}
    \int d\mathbf{r} Y_{l^{\prime}m^{\prime}}^{\ast}(\mathbf{r}) Y_{lm}(\mathbf{r}) = \delta_{l^{\prime}l} \delta_{m^{\prime}m}. \label{equ:sphericalharmonicorthogonal}
\end{equation}
This relation separates the radial integrals from the spin-angular part, so that the latter can be handled separately through standard angular-momentum algebra.
For matrix elements $\langle (s_i^{\prime},s_j^{\prime},S^{\prime},L^{\prime}) |\hat{O}| (s_i,s_j,S,L) \rangle_{J}$ of the spin-dependent operators $\hat{O}$, we use
\begin{align}
    &\langle \mathbf{S}_i \cdot \mathbf{S}_j \rangle
    = \frac{1}{2} \big(S(S+1)-s_i(s_i+1)-s_j(s_j+1)\big)
    \delta_{s_i s_i^{\prime}}\delta_{s_j s_j^{\prime}}\delta_{S S^{\prime}}\delta_{L L^{\prime}}, \label{equ:CasimirSS} \\
    &\langle \mathbf{L} \cdot \mathbf{S}_i \rangle
    = (-1)^{S+L^{\prime}+J} (-1)^{s_i^{\prime}+s_j^{\prime}+S+1}
    \sqrt{(2S^{\prime}+1)(2S+1)} \sqrt{(2L+1)(L+1)L} \sqrt{(2s_i+1)(s_i+1)s_i} \nonumber \\
    &\qquad \times
    \begin{Bmatrix}
        S^{\prime} & S & 1 \\
        L & L^{\prime} & J
    \end{Bmatrix}
    \begin{Bmatrix}
        s_i & s_i^{\prime} & 1 \\
        S^{\prime} & S & s_j^{\prime}
    \end{Bmatrix}
    \delta_{s_i s_i^{\prime}}\delta_{s_j s_j^{\prime}}\delta_{L L^{\prime}}, \label{equ:CasimirLS1} \\
    &\langle \mathbf{L} \cdot \mathbf{S}_j \rangle
    = (-1)^{S+L^{\prime}+J} (-1)^{s_i+s_j+S^{\prime}+1}
    \sqrt{(2S^{\prime}+1)(2S+1)} \sqrt{(2L+1)(L+1)L} \sqrt{(2s_j+1)(s_j+1)s_j} \nonumber \\
    &\qquad \times
    \begin{Bmatrix}
        S^{\prime} & S & 1 \\
        L & L^{\prime} & J
    \end{Bmatrix}
    \begin{Bmatrix}
        s_j & s_j^{\prime} & 1 \\
        S^{\prime} & S & s_i^{\prime}
    \end{Bmatrix}
    \delta_{s_i s_i^{\prime}}\delta_{s_j s_j^{\prime}}\delta_{L L^{\prime}}, \label{equ:CasimirLS2} \\
    &\langle \frac{3(\mathbf{S}_i\cdot\mathbf{r}_{ij}) (\mathbf{S}_j\cdot\mathbf{r}_{ij})}{r_{ij}^2} - \mathbf{S}_i\cdot\mathbf{S}_j \rangle
    = (-1)^{S+L^{\prime}+J} (-1)^{L^{\prime}} \sqrt{30} \sqrt{(2S^{\prime}+1)(2S+1)} \sqrt{(2L^{\prime}+1)(2L+1)} \nonumber \\
    &\qquad \times \sqrt{(2s_i+1)(s_i+1)s_i} \sqrt{(2s_j+1)(s_j+1)s_j}
    \begin{pmatrix}
        L^{\prime} & 2 & L \\
        0 & 0 & 0
    \end{pmatrix}
    \begin{Bmatrix}
        S^{\prime} & S & 2 \\
        L & L^{\prime} & J
    \end{Bmatrix}
    \begin{Bmatrix}
        s_i^{\prime} & s_i & 1 \\
        s_j^{\prime} & s_j & 1 \\
        S^{\prime} & S & 2
    \end{Bmatrix}
    \delta_{s_i s_i^{\prime}}\delta_{s_j s_j^{\prime}}, \label{equ:CasimirTensor}
\end{align}
These equations are evaluated channel by channel and supply the precomputed spin-coupling coefficients multiplying the potential terms of Eqs.~(\ref{equ:Vijcont})--(\ref{equ:Vijtens}). 

An additional complication of the GI framework is that different operators are most naturally treated in different representations. 
Some are evaluated in coordinate space, whereas others, especially the kinetic-energy term and smearing factors, are more naturally handled in momentum space.
The calculation is therefore carried out in a mixed representation. For example, the kinetic-energy term is most conveniently evaluated in momentum space:
\begin{equation}
    \langle \phi_{m}|\sqrt{p^2+m^2} |\phi_{n}\rangle = \int p^2 dp R_{ml}^{p} \sqrt{p^2+m^2} R_{nl}^{p}. \label{equ:kineticenergymatrixelement}
\end{equation}
For the potential operators, one may insert two sets of complete bases to handle the mixed representation as needed. 
In particular, matrix elements of an operator in the form $\hat{A}(p)\hat{B}(r)\hat{A}(p)$ are written as
\begin{equation}
    \langle \phi_{m}|\hat{A}(p)\hat{B}(r)\hat{A}(p)|\phi_{n}\rangle
    = \sum_{i} \sum_{j} \langle \phi_{m}|\hat{A}(p) |\phi_{i}\rangle\langle\phi_{i}| \hat{B}(r) |\phi_{j}\rangle\langle\phi_{j}| \hat{A}(p)|\phi_{n}\rangle, \label{equ:potentialmatrixelement}
\end{equation}
where
\begin{align}
    \langle \phi_{m}|\hat{A}(p) |\phi_{i}\rangle
    &= \int p^2 dp R_{ml}^{p} \hat{A}(p) R_{il}^{p}, \label{equ:Apmatrixelement} \\
    \langle\phi_{i}| \hat{B}(r) |\phi_{j}\rangle
    &= \int r^2 dr R_{il}^{r} \hat{B}(r) R_{jl}^{r}, \label{equ:Brmatrixelement} \\
    \langle\phi_{j}| \hat{A}(p)|\phi_{n}\rangle
    &= \int p^2 dp R_{jl}^{p} \hat{A}(p) R_{nl}^{p}. \label{equ:Apnmatrixelement}
\end{align}
This representation is numerically advantageous as it separates the matrix element into simple coordinate and momentum space components, each of which can be evaluated independently. 
Since the Gaussian basis is not orthogonal, these matrices must be transformed into a stable orthonormal representation before entering the eigenvalue problem. 
In the present analysis, this approach is carried out through an auxiliary random-matrix construction, as described in the next subsection. 
Once the eigenvectors are obtained, the same transformed basis is used to evaluate the root-mean-square (RMS) radius.

\subsection{Random-Matrix Construction for Orthogonal Basis}
\label{sec:cholesky}

A practical difficulty is that, in a nonorthogonal basis, the matrices in Eqs.~(\ref{equ:Apmatrixelement})--(\ref{equ:Apnmatrixelement}) cannot be combined by multiplication as in Eq.~(\ref{equ:potentialmatrixelement}).
As a result, the Hamiltonian must first be expressed in an orthogonal basis before standard diagonalization can be carried out.
To overcome this difficulty, we construct an orthogonal basis by means of an auxiliary random-matrix method. 
The resulting basis allows the operator matrices to be combined consistently and the Hamiltonian to be diagonalized in a standard eigenvalue problem.

Concretely, a random matrix $\mathbf{X}$ is generated with elements uniformly distributed in the interval $[-1,1]$, and then symmetrized according to
\begin{equation}
    \mathbf{R} = \mathbf{X} + \mathbf{X}^{T}.
\end{equation}
The generalized eigenvalue problem
\begin{equation}
    \mathbf{R}\mathbf{v}_{k} = \lambda_{k}\mathbf{N}\mathbf{v}_{k}
\end{equation}
is then solved to produce the eigenvalues $\lambda_k$ and eigenvectors $\mathbf{v}_k$. 
Because $\mathbf{R}$ is symmetric and $\mathbf{N}$ is real symmetric positive definite, the resulting eigenvectors are mutually orthogonal with respect to the overlap metric,
which can be expressed as
\begin{equation}
    \mathbf{v}_{i}^{T}\mathbf{N}\mathbf{v}_{j} = 0 \qquad (i \neq j).
    \label{equ:standardEP}
\end{equation}
Therefore, the random matrix $\mathbf{R}$ serves only as a device for selecting a complete $\mathbf{N}$-orthogonal basis from the original Gaussian subspace. 
Different random seeds may rotate this basis, but they do not alter the underlying variational space from which the physical spectrum is extracted.

Collecting these eigenvectors $\mathbf{v}_k$ into a matrix $\mathbf{V}$ whose rows contain the transformed basis vectors, 
one can normalize them with respect to the overlap matrix by evaluating the norm factors
\begin{equation}
    n_{k} = \sqrt{|\mathbf{v}_{k}^{T}\mathbf{N}\mathbf{v}_{k}|},
\end{equation}
and dividing each row by the corresponding factor $n_k$. After this step the transformed basis satisfies
\begin{equation}
    \mathbf{V}\mathbf{N}\mathbf{V}^{T} = \mathbf{I}
\end{equation}
up to numerical precision. All precomputed operator matrices in Eqs.~(\ref{equ:Apmatrixelement})--(\ref{equ:Apnmatrixelement}) are then mapped into this orthogonal basis through the transformation
\begin{equation}
    \mathbf{M}^{\prime} = \mathbf{V}\mathbf{M}\mathbf{V}^{T},
\end{equation}
from which the transformed kinetic energy and potentials are assembled to construct the full Hamiltonian matrix $\mathbf{H}^{\prime}$. Since the overlap matrix has now been reduced to the identity, the mass spectrum is obtained by solving the standard eigenvalue problem
\begin{equation}
    \mathbf{H}^{\prime}\mathbf{u} = E\mathbf{u}.
\end{equation}
The final coefficient vectors in the original Gaussian basis are recovered by back-transformation 
\begin{equation}
    \mathbf{C} = \mathbf{U}\mathbf{V},
\end{equation}
where the rows of $\mathbf{U}$ are the eigenvectors $\mathbf{u}$ of the orthogonalized Hamiltonian $\mathbf{H}^{\prime}$.

\begin{figure}[!htbp]
    \centering
    \subfigure[$\ n=1$]{\includegraphics[width=0.31\textwidth]{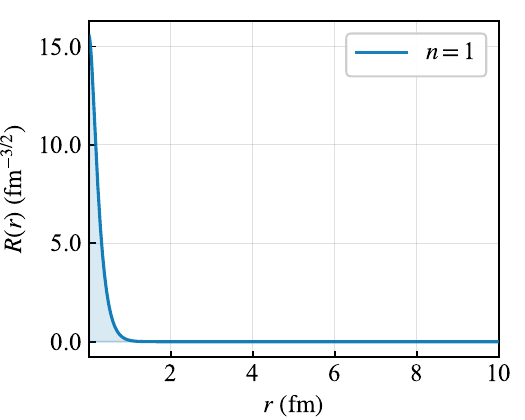}}
    \subfigure[$\ n=2$]{\includegraphics[width=0.31\textwidth]{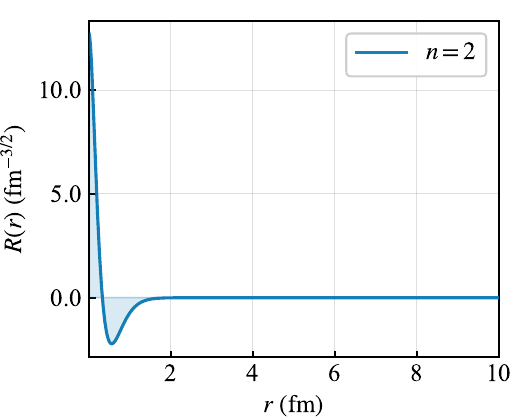}}
    \subfigure[$\ n=3$]{\includegraphics[width=0.31\textwidth]{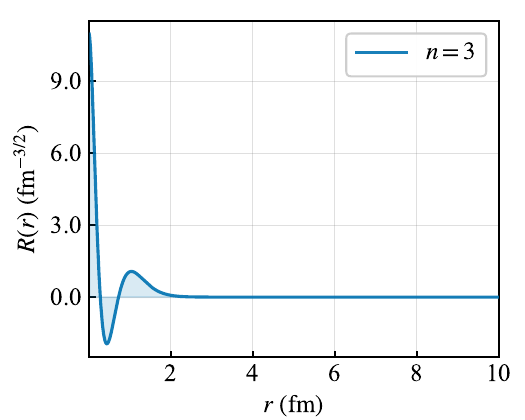}}\\
    \subfigure[$\ n=4$]{\includegraphics[width=0.31\textwidth]{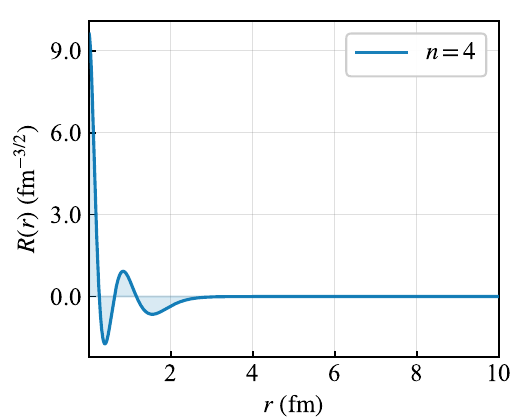}}
    \subfigure[$\ n=5$]{\includegraphics[width=0.31\textwidth]{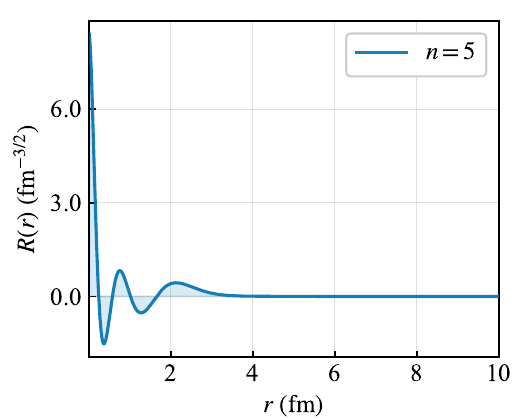}}
    \subfigure[$\ n=6$]{\includegraphics[width=0.31\textwidth]{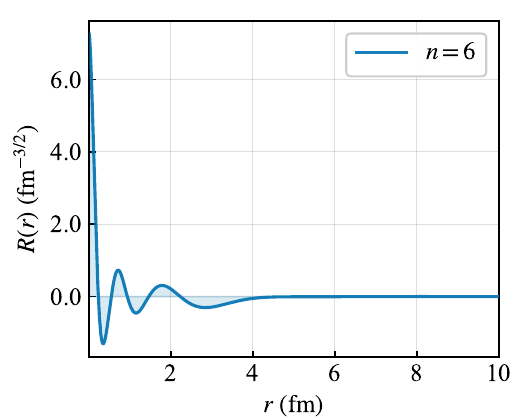}}\\
    \subfigure[$\ n=7$]{\includegraphics[width=0.31\textwidth]{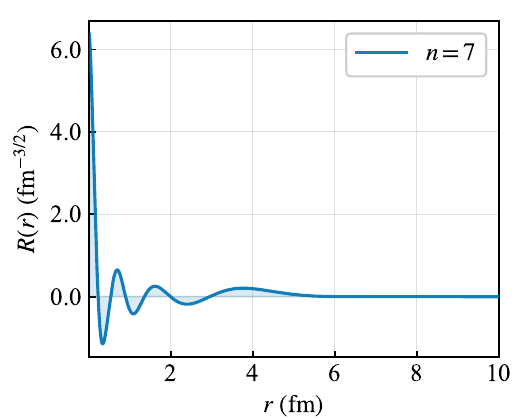}}
    \subfigure[$\ n=8$]{\includegraphics[width=0.31\textwidth]{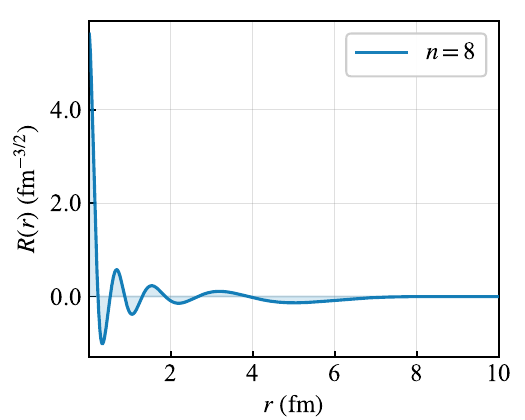}}
    \subfigure[$\ n=9$]{\includegraphics[width=0.31\textwidth]{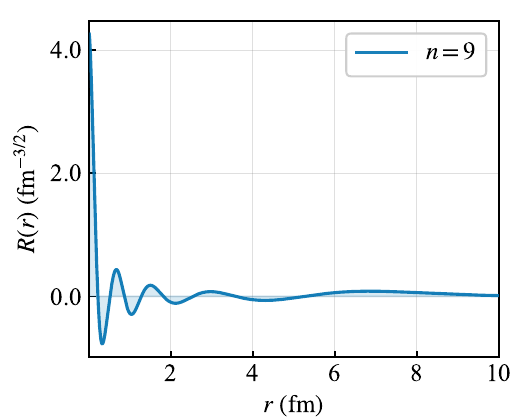}}
    \caption{Radial wave functions for the $n{}^{3}S_{1}$ charmonium states obtained from the random-matrix orthogonalization procedure. 
    The number of nodes increases consistently with the radial quantum number $n$, showing that the orthogonal basis preserves the expected multi-node structure of the excitations. 
    The absence of strong oscillations in the higher excitations supports the numerical reliability of the orthogonalization approach.}
    \label{fig:charmonium-wavefunctions}
\end{figure}

In practical calculations, this method produces radial wave functions with the expected multi-node structure for highly excited states, while retaining numerical stability in the constructed Hamiltonian. 
As shown in Fig.~\ref{fig:charmonium-wavefunctions}, the wave functions of $n{}^{3}S_{1}$ charmonium exhibit the systematic increase in the number of nodes with radial excitation, 
which confirms that the orthogonalization preserves the physical structure of the solutions. 
The smooth progression of these node patterns also indicates that the random-matrix construction yields a reliable orthogonal basis for the calculation.

It should be noted that the coefficient vectors $\mathbf{C}$ do not have unit norm in the ordinary Euclidean sense. 
We also examined a Cholesky-based orthogonalization scheme, which yields coefficient vectors with unit norm more directly. 
However, the corresponding wave functions do not exhibit the expected oscillatory and nodal structure. 
For this reason, the random-matrix construction was adopted in the final implementation. 
If the radial wave function is written in terms of Eq.~(\ref{equ:GEMbasis}) as
\begin{equation}
    \psi_{nl} = \sum_{i=1}^{N_{\rm max}} C_{ni} \phi_{il}, \label{equ:wavefunctionC}
\end{equation}
with radial number $n$ and angular momentum $l$, the RMS radius is evaluated from the normalized expectation value
\begin{equation}
    r^{\rm rms}_{n} = \sqrt{\frac{\langle \psi_{nl} | r^2 | \psi_{nl} \rangle}{\langle \psi_{nl} | \psi_{nl} \rangle}}
    = \left(\frac{\sum_{ij} C_{ni} C_{nj} \langle \phi_{il} | r^2 | \phi_{jl} \rangle}{\sum_{ij} C_{ni} C_{nj} \langle \phi_{il} | \phi_{jl} \rangle}\right)^{\frac{1}{2}}, \label{equ:rmsExpectation}
\end{equation}
which, in matrix form, is written as
\begin{equation}
    r^{\rm rms}_{n} = \left(\frac{\mathbf{C}_{n}\mathbf{R}^{2}\mathbf{C}_{n}^{T}}{\mathbf{C}_{n}\mathbf{N}\mathbf{C}_{n}^{T}}\right)^{\frac{1}{2}}. \label{equ:rmsMatrix}
\end{equation}
Here $\mathbf{R}^{2}$ denotes the matrix representation of the operator $r^{2}$ and $\mathbf{N}$ is the corresponding overlap matrix. 
In the final implementation, the normalized expectation value is converted from natural units to fm.

\subsection{Gaussian Quadrature Implementation}

The matrix elements entering the GEM calculation reduce to a large set of integrals over either $r\in[0,\infty)$ or $p\in[0,\infty)$. 
To compute these integrals, we introduce a dedicated Gaussian quadrature for the weight function
\begin{equation}
    \rho(x) = e^{-x^{2}}, \qquad x \in [0,\infty),
\end{equation}
which is the natural weight appearing after the products of Gaussian basis functions.

To construct the Gaussian quadrature adapted to the weight function $\rho(x)=e^{-x^{2}}$, 
one first needs a family of orthogonal polynomials whose roots can be used as the quadrature nodes. 
The orthogonal polynomials are obtained by applying the Gram--Schmidt procedure to the monomials $1,x,x^2,\ldots$ with respect to the inner product
\begin{equation}
    (f,g) = \int_{0}^{\infty} f(x)g(x)\rho(x)\,dx.
\end{equation}
In practice, one first constructs a monic family of orthogonal polynomials in the form
\begin{equation}
    \widetilde{\psi}_n(x)=x^n+\sum_{k=0}^{n-1} c_{nk}\widetilde{\psi}_k(x),
\end{equation}
with coefficients determined by the orthogonality condition
\begin{equation}
    c_{nk}=-\frac{(x^n,\widetilde{\psi}_k)}{(\widetilde{\psi}_k,\widetilde{\psi}_k)}.
\end{equation}
The resulting polynomials are then normalized according to
\begin{equation}
    \psi_n(x)=\frac{\widetilde{\psi}_n(x)}{\sqrt{(\widetilde{\psi}_n,\widetilde{\psi}_n)}}.
\end{equation}
This construction avoids numerical differentiation and gives direct control over the polynomial coefficients at arbitrary precision.

For a quadrature of order $n$, the nodes $x_k$ are chosen as the $n$ roots of the polynomial $\psi_n(x)$. 
Once these nodes are determined, the corresponding weights $A_k$ are obtained from exact moment matching,
\begin{equation}
    \sum_{k=1}^{n} A_k x_k^m = \mu_m, \qquad m=0,1,\ldots,n-1,
\end{equation}
where $\mu_m$ are the moments of the weight function defined by
\begin{equation}
    \mu_m = \int_{0}^{\infty} x^{m} \rho(x)\,dx.
\end{equation}
Solving this linear system yields the quadrature rule
\begin{equation}
    \int_{0}^{\infty} h(x) \rho(x)\,dx \approx \sum_{k=1}^{n} A_k h(x_k).
\end{equation}
With this choice of nodes $x_k$ and weights $A_k$, the quadrature rule exactly reproduces the integral whenever $h(x)$ is a polynomial of degree no greater than $2n-1$. 
This is the standard optimality property of an \(n\)-point Gaussian quadrature rule.

This quadrature rule can be verified in several forms relevant to the present calculation. First, one checks the weighted integral against analytic examples such as
\begin{equation}
    \int_{0}^{\infty} \sin x\, e^{-x^{2}} dx \approx \sum_{k=1}^{n} A_k \sin(x_k).
\end{equation}
Second, the scaling implementation is required for general Gaussian factor $e^{-t x^2}$. With the substitution $y=\sqrt{t}\,x$, one obtains
\begin{equation}
    \int_{0}^{\infty} h(x)e^{-t x^{2}} dx
    = \frac{1}{\sqrt{t}} \int_{0}^{\infty} h\!\left(\frac{y}{\sqrt{t}}\right)e^{-y^{2}}dy
    \approx \frac{1}{\sqrt{t}}\sum_{k=1}^{n} A_k h\!\left(\frac{x_k}{\sqrt{t}}\right).
\end{equation}
This scaling relation is repeatedly used for products of two Gaussian basis functions when evaluating matrix elements.
The same logic applies to radial integrals for the Gaussian basis (\ref{equ:GEMbasis}). In practice, after combining two basis functions one encounters integrals of the form
\begin{equation}
    \int_{0}^{\infty} h(r)e^{-2\nu r^{2}} r^{2}dr,
\end{equation}
which are evaluated through the rescaling $y=\sqrt{2\nu}\,r$ as
\begin{equation}
    \int_{0}^{\infty} h(r)e^{-2\nu r^{2}} r^{2}dr
    \approx \frac{1}{\sqrt{2\nu}} \sum_{k=1}^{n} A_k
    h\!\left(\frac{x_k}{\sqrt{2\nu}}\right)
    \left(\frac{x_k}{\sqrt{2\nu}}\right)^{2}.
\end{equation}
This is precisely the form needed for normalized Gaussian orbitals, where the quadrature reproduces the normalization factor to numerical precision.

In practical calculations, the nodes and weights are precomputed at sufficiently high order (usually $n\sim50$) and reused for the matrix-element evaluation. 
In this way, the matrix elements are computed with a fixed quadrature rule for the Gaussian basis, instead of relying on adaptive integrations. 
This makes the assembly of the overlap and Hamiltonian matrices much more efficient, while preserving the numerical precision needed for the spectroscopy analysis.

\section{Parameter Determination}
\label{sec:fit}

This section explains how the parameters of the screen-modified GI model are determined before the spectrum analysis. 
We first describe the fitting targets, tools, strategy, and failure modes used to obtain stable parameter sets, 
and then summarize the final fitted parameters with the resulting quality of the charmonium and bottomonium spectra.

\subsection{Fit Strategy and Constraints}

The fitting target consists of the experimentally established heavy mesons used as calibration inputs for the model. 
For charmonium and bottomonium, the reference states are the available candidates including $nS$, $nP$, and $nD$ families. 
At the same time, the fit is designed to determine the model parameters introduced in Sec.~\ref{sec:model}, containing the constituent quark masses, 
the constant term $c$, the smearing parameters $\sigma_0$ and $s$, the confinement parameters $b$ and $\mu$, 
and the relativistic exponents $\epsilon_{\rm Coul}$, $\epsilon_{\rm cont}$, $\epsilon_{\rm so(\nu)}$, $\epsilon_{\rm so(s)}$, and $\epsilon_{\rm tens}$. 
Particular attention is paid to the screening parameter $\mu$, since its inverse may be interpreted as string-breaking distance of the screened confining interaction. 
During the fit this scale is required to remain consistent with the lattice-QCD studies of string breaking phenomenon.

The fitting tools combine the numerical Hamiltonian solver with a standard $\chi^2$ minimization. The function is written as
\begin{equation}
    \chi^{2} = \sum_{i} \left(\frac{M_{i}^{\rm th}-M_{i}^{\rm exp}}{M_i^{\rm err}}\right)^{2},
\end{equation}
where $M_{i}^{\rm th}$ and $M_{i}^{\rm exp}$ denote the calculated and experimental masses of the reference states, 
and $M_i^{\rm err}$ is the experimental uncertainty for each input state. 
The minimization is carried out with the \texttt{ROOT::Minuit2} library using the \texttt{Migrad} algorithm, 
while the theoretical masses entering $\chi^2$ are computed by the \texttt{gemstore} program developed for the present analysis. 
In practice, each parameter set is passed to the GEM solver, the Hamiltonian is diagonalized, and the resulting spectrum is returned to the minimizer for the next update.

The fitting strategy proceeds in several steps. First, the Gaussian basis parameters such as $N_{\rm max}=20$, $r_{\rm max}=30$ fm, and $r_{\rm min}=0.1$ fm are selected through convergence tests, 
so that the numerical basis is sufficiently stable before the parameter optimization. Second, a unified fit to the established heavy-meson systems, 
including charmonium, bottomonium, $B_c$, $D_s$, $B_s$, $D$, and $B$ mesons, is performed to constrain the system-independent parameters, 
namely the quark masses, $c$, $\sigma_0$, and $s$.
Because this stage is comparatively high dimensional, several random initializations are launched with \texttt{Minuit2/Migrad} and the solution of minimal $\chi^2$ is retained as the global optimum.
Third, with those unified parameters fixed, the charmonium and bottomonium calibration states in Tables~\ref{tab:sm-charmonium-comparison} and \ref{tab:sm-bottomonium-comparison} are used to determine the system-dependent parameters $b$, $\mu$, 
and the relativistic exponents for each sector separately.
With the system-independent quantities held fixed the second-stage minimum is unique: repeated runs from the staged starting point return the same production parameters.
Finally, the fitting results are examined through their $\chi^2$, the extracted screening scale $\mu$, 
the onset of the saturated spectrum, and the behavior of the corresponding radial wave functions.
Fit quality is summarized by residual statistics of $\delta M_i=M_i^{\rm th}-M_i^{\rm exp}$ on the calibration sets, reported as descriptive references rather than as propagated errors on $M_\infty$.
Over the $17$ charmonium and $19$ bottomonium states one finds
${\rm RMS}(\delta M)=34$ and $24\,\mathrm{MeV}$,
sample standard deviations $\sigma_{\delta M}=34$ and $23\,\mathrm{MeV}$,
and mean absolute residuals $20$ and $11\,\mathrm{MeV}$, respectively.
The dominant outliers are higher vector states above open-flavor thresholds; restricted to the better-controlled low-lying subsets the RMS residuals fall to about $10\,\mathrm{MeV}$ ($c\bar{c}$) and $4\,\mathrm{MeV}$ ($b\bar{b}$).
Formal \texttt{Minuit2} parameter errors are not used: with experimental $M_{\rm err}$ alone in $\chi^2$, the Hessian underestimates the true model uncertainty, so the production parameters and $M_\infty$ are quoted as central values only.

Several failure modes are also monitored during the fit. First, if one enforces a fully unified fit in which charmonium and bottomonium share the same set of parameters, 
the charmonium limitation scale is shifted to about 4.83 GeV, while the bottomonium limitation scale is shifted to about 11.52 GeV. 
Second, if the system-independent parameters such as $c$, $\sigma_0$, and $s$ are not fixed in the separated stage, 
the charmonium string-breaking distance is reduced to about 1.19 fm and the corresponding limitation scale is shifted down to about 4.62 GeV, which is obviously incorrect. 
Third, if one fits only the charmonium spectrum individually without fixing any parameter, the resulting string-breaking distance is about 1.25 fm, which is acceptable by itself, 
but the corresponding limitation scale is still shifted to about 4.61 GeV. 
Finally, if one chooses $N_{\rm max}=16$ or an even smaller value, the numerical precision is reduced because of basis saturation; as a result, 
the radial wave functions of the highly excited states oscillate strongly and develop excessively wide tails.

\subsection{Final Parameters and Fit Quality}

The final fitted parameters are summarized in Table~\ref{tab:GIScreen-parameter}. The entries in the last column record the parameters obtained in the unified fit to heavy mesons.
Starting from this unified solution, the charmonium and bottomonium sectors are refitted separately to determine the system-dependent parameters, such as $b$, $\mu$, and the relativistic exponents. 
The quality of the final fit is demonstrated by the spectra in Tables~\ref{tab:sm-charmonium-comparison} and \ref{tab:sm-bottomonium-comparison}, 
where the states directly test how well the fitted parameters reproduce the quarkonium spectrum.
In these tables, $M_{\rm err}$ denotes the experimental mass uncertainty for each state, given in MeV and rounded up to the nearest integer.

More specifically, the fit quality is best for the low-lying states, including the $1S$, $1P$, and lowest $D$-wave levels in both charmonium and bottomonium, 
for which the mass deviations are typically only a few MeV. In the charmonium sector, states such as $\eta_c(1S)$, $J/\psi$, $h_c(1P)$, the $\chi_{cJ}(1P)$ multiplet, 
$\psi(3770)$, $\psi_2(3823)$, and $\psi_3(3842)$ are reproduced well, and even the $\psi(5S)$ assignment remains close to the observed mass of $\psi(4415)$. 
The more visible failures appear for the higher vector excitations, especially the states identified here with $2^3D_1$ and $3^3D_1$, 
whose predicted masses lie below the experimental candidates $\psi(4160)$ and $\psi(4360)$. 
In the bottomonium sector the agreement is similarly good for the low-lying $S$-wave, $P$-wave, and $1D$ states, 
whereas the largest errors appear for the higher vector states $\Upsilon(4S)$, $\Upsilon(10860)$, and $\Upsilon(11020)$.

This pattern is physically reasonable. The well-reproduced states are mostly low-lying conventional quarkonium states and lie in the region where the potential model remains reliable. 
By contrast, the states with the largest errors are those near or above the open-flavor thresholds. The first and most important reason is the intrinsic limitation of the screened potential model: 
although the screened interaction is intended to mimic coupled-channel effects, it cannot replace the treatment of open-flavor channels. 
The second reason is the absence of the $S$-$D$ wave mixing mechanism, which is expected to be relevant for higher vector states. 
The third reason is the uncertainty of the experimental candidates, because some of them may correspond to different $n^{2S+1}L_J$ assignments or may even contain exotic components. 
The mismatches in Tables~\ref{tab:sm-charmonium-comparison} and \ref{tab:sm-bottomonium-comparison} should therefore be understood mainly as limitations of the model for unquenched effects and open-flavor channels.

\renewcommand{\tabcolsep}{0.20cm}
\renewcommand{\arraystretch}{1.1}
\begin{table*}[!htbp]
\caption{Final parameters of the screen-modified GI model. The last column lists the parameters from the unified heavy-meson fit, 
    while the charmonium and bottomonium columns give the final production values after the system-dependent refits.}
	\label{tab:GIScreen-parameter}
	\begin{tabular}{lccc}
		\toprule[1.5pt]\toprule[0.5pt]
		Parameters & Charmonium & Bottomonium & Heavy meson \\
		\midrule[0.5pt]
        $m_n\ ({\rm GeV})$          & \multicolumn{2}{c}{\dots} & 0.456 \\
        $m_s\ ({\rm GeV})$          & \multicolumn{2}{c}{\dots} & 0.617 \\
        $m_c\ ({\rm GeV})$          & \multicolumn{2}{c}{1.805} & 1.805 \\
        $m_b\ ({\rm GeV})$          & \multicolumn{2}{c}{5.151} & 5.151 \\
        $c\ ({\rm GeV})$            & \multicolumn{2}{c}{-0.648} & -0.648 \\
        $\sigma_0\ ({\rm GeV})$     & \multicolumn{2}{c}{1.771} & 1.771 \\
        $s$                         & \multicolumn{2}{c}{1.146} & 1.146 \\
        $b\ ({\rm GeV}^2)$          & 0.256 & 0.247 & 0.252 \\
        $\mu\ ({\rm GeV})$          & 0.144 & 0.122 & 0.135 \\
        $\epsilon_{\rm Coul}$       & 0.0 & 0.0 & 0.0 \\
        $\epsilon_{\rm cont}$       & -0.359 & -0.498 & -0.320 \\
        $\epsilon_{\rm so(\nu)}$   & -0.499 & -0.146 & -0.343 \\
        $\epsilon_{\rm so(s)}$      & 1.000 & -0.502 & 1.000 \\
        $\epsilon_{\rm tens}$       & -0.500 & -0.846 & -0.500 \\
		\bottomrule[0.5pt]\bottomrule[1.5pt]
	\end{tabular}
\end{table*}

\renewcommand{\tabcolsep}{0.20cm}
\renewcommand{\arraystretch}{1.1}
\begin{table*}[!htbp]
	\caption{Comparison between the calculated and experimental charmonium spectrum. 
    Listed are the spectroscopic assignment $n^{2S+1}L_J$, quantum numbers $J^{PC}$, calculated mass $M_{\rm th}$ (in MeV), experimental mass $M_{\rm exp}$ (in MeV), 
    adopted experimental uncertainty $M_{\rm err}$ (in MeV), and the corresponding state assignment.}
	\label{tab:sm-charmonium-comparison}
	\begin{tabular}{ccrrcl}
		\toprule[1.5pt]\toprule[0.5pt]
		$n^{2S+1}L_{J}$ & $J^{PC}$ & $M_{\rm th}$ & $M_{\rm exp}$ & $M_{\rm err}$ & State \\
		\midrule[0.5pt]
		$1^{1}S_{0}$ & $0^{-+}$    & 2990    & 2984 & 1    & $\eta_c(1S)$ \\
		$2^{1}S_{0}$ & $0^{-+}$    & 3627    & 3638 & 1    & $\eta_c(2S)$ \\
		$1^{3}S_{1}$ & $1^{--}$    & 3103    & 3097 & 1    & $J/\psi(1S)$ \\
		$2^{3}S_{1}$ & $1^{--}$    & 3672    & 3686 & 1    & $\psi(2S)$ \\
		$3^{3}S_{1}$ & $1^{--}$    & 4015    & 4040 & 4    & $\psi(4040)$ \\
		$4^{3}S_{1}$ & $1^{--}$    & 4252    & 4222 & 3    & $\psi(4230)$ \\
		$5^{3}S_{1}$ & $1^{--}$    & 4422    & 4415 & 5    & $\psi(4415)$ \\
		$1^{1}P_{1}$ & $1^{+-}$    & 3522    & 3525 & 1    & $h_c(1P)$ \\
		$1^{3}P_{0}$ & $0^{++}$    & 3419    & 3415 & 1    & $\chi_{c0}(1P)$ \\
		$1^{3}P_{1}$ & $1^{++}$    & 3506    & 3511 & 1    & $\chi_{c1}(1P)$ \\
		$1^{3}P_{2}$ & $2^{++}$    & 3563    & 3556 & 1    & $\chi_{c2}(1P)$ \\
		$2^{3}P_{2}$ & $2^{++}$    & 3935    & 3923 & 1    & $\chi_{c2}(3930)$ \\
		$1^{3}D_{1}$ & $1^{--}$    & 3787    & 3774 & 1    & $\psi(3770)$ \\
		$2^{3}D_{1}$ & $1^{--}$    & 4083    & 4191 & 5    & $\psi(4160)$ \\
		$3^{3}D_{1}$ & $1^{--}$    & 4296    & 4374 & 7    & $\psi(4360)$ \\
		$1^{3}D_{2}$ & $2^{--}$    & 3819    & 3824 & 1    & $\psi_2(3823)$ \\
		$1^{3}D_{3}$ & $3^{--}$    & 3843    & 3843 & 1    & $\psi_3(3842)$ \\
		\bottomrule[0.5pt]\bottomrule[1.5pt]
	\end{tabular}
\end{table*}

\renewcommand{\tabcolsep}{0.20cm}
\renewcommand{\arraystretch}{1.1}
\begin{table*}[!htbp]
	\caption{Comparison between the calculated and experimental bottomonium spectrum. 
    Listed are the spectroscopic assignment $n^{2S+1}L_J$, quantum numbers $J^{PC}$, calculated mass $M_{\rm th}$ (in MeV), experimental mass $M_{\rm exp}$ (in MeV), 
    adopted experimental uncertainty $M_{\rm err}$ (in MeV), and the corresponding state assignment.}
	\label{tab:sm-bottomonium-comparison}
	\begin{tabular}{ccrrcl}
		\toprule[1.5pt]\toprule[0.5pt]
		$n^{2S+1}L_{J}$ & $J^{PC}$ & $M_{\rm th}$ & $M_{\rm exp}$ & $M_{\rm err}$ & State \\
		\midrule[0.5pt]
		$1^{1}S_{0}$ & $0^{-+}$    & 9400     & 9399  & 2    & $\eta_b(1S)$ \\
		$2^{1}S_{0}$ & $0^{-+}$    & 9996     & 9999  & 4    & $\eta_b(2S)$ \\
		$1^{3}S_{1}$ & $1^{--}$    & 9458     & 9460  & 1    & $\Upsilon(1S)$ \\
		$2^{3}S_{1}$ & $1^{--}$    & 10021    & 10023 & 1    & $\Upsilon(2S)$ \\
		$3^{3}S_{1}$ & $1^{--}$    & 10358    & 10355 & 1    & $\Upsilon(3S)$ \\
		$4^{3}S_{1}$ & $1^{--}$    & 10604    & 10579 & 2    & $\Upsilon(4S)$ \\
		$5^{3}S_{1}$ & $1^{--}$    & 10797    & 10885 & 3    & $\Upsilon(10860)$ \\
		$6^{3}S_{1}$ & $1^{--}$    & 10958    & 11000 & 4    & $\Upsilon(11020)$ \\
		$1^{1}P_{1}$ & $1^{+-}$    & 9892     & 9899  & 1    & $h_b(1P)$ \\
		$2^{1}P_{1}$ & $1^{+-}$    & 10259    & 10260 & 2    & $h_b(2P)$ \\
		$1^{3}P_{0}$ & $0^{++}$    & 9856     & 9859  & 1    & $\chi_{b0}(1P)$ \\
		$2^{3}P_{0}$ & $0^{++}$    & 10236    & 10233 & 1    & $\chi_{b0}(2P)$ \\
		$1^{3}P_{1}$ & $1^{++}$    & 9886     & 9893  & 1    & $\chi_{b1}(1P)$ \\
		$2^{3}P_{1}$ & $1^{++}$    & 10256    & 10256 & 1    & $\chi_{b1}(2P)$ \\
		$3^{3}P_{1}$ & $1^{++}$    & 10522    & 10513 & 1    & $\chi_{b1}(3P)$ \\
		$1^{3}P_{2}$ & $2^{++}$    & 9907     & 9912  & 1    & $\chi_{b2}(1P)$ \\
		$2^{3}P_{2}$ & $2^{++}$    & 10270    & 10269 & 1    & $\chi_{b2}(2P)$ \\
		$3^{3}P_{2}$ & $2^{++}$    & 10532    & 10524 & 1    & $\chi_{b2}(3P)$ \\
		$1^{1}D_{2}$ & $2^{-+}$    & 10165    & 10164 & 2    & $\Upsilon_2(1D)$ \\
		\bottomrule[0.5pt]\bottomrule[1.5pt]
	\end{tabular}
\end{table*}

\section{Spectrum Results and Limitation Analysis}
\label{sec:results}

This section presents the extended spectral results obtained with the final parameters and then discusses the limitation analysis. 
We first summarize the calculated masses and root-mean-square radii, together with their graphical representations and possible state assignments, 
and then turn to a separate discussion of the critical radial quantum number at which the limitation sets in.

\subsection{Extended Spectra Results}

The calculated masses and root-mean-square radii are presented in Tables~\ref{tab:charmonium-spectra} and \ref{tab:bottomonium-spectra} for the extended charmonium and bottomonium spectra. 
A unified pattern is seen in all channels. As the radial quantum number $n$ increases, the masses gradually approach a saturation region, 
whereas the RMS radii increase rapidly and tend toward a divergent behavior. This shows that the highly excited states become increasingly extended in configuration space, 
even though the corresponding masses change only weakly. In this sense, the tables already indicate the limitation scale of mass convergence and mass-radius decoupling. 
Here we list states only up to $n=14$, although the basis number is taken as $N_{\rm max}=20$, in order to exclude unphysical states 
that are affected by basis saturation and therefore no longer provide reliable spectral information.

\begin{figure*}[!htbp]
    \centering
    \subfigure[\(\ \)Charmonium energy levels.]{\includegraphics[width=0.47\textwidth]{charmonium_energy_levels.pdf}}
    \subfigure[\(\ \)Bottomonium energy levels.]{\includegraphics[width=0.47\textwidth]{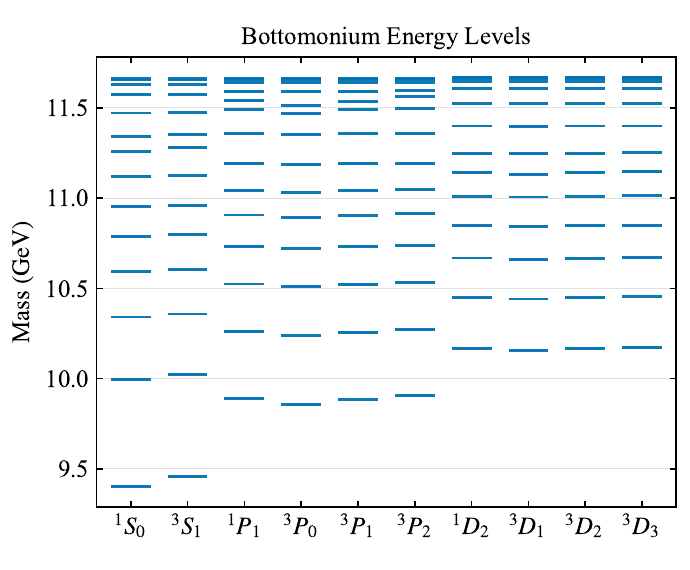}}
    \caption{Calculated energy levels for the extended charmonium and bottomonium spectra. 
    The two panels visualize the mass values listed in Tables~\ref{tab:charmonium-spectra} and \ref{tab:bottomonium-spectra}. 
    The low-lying states remain clearly separated, whereas the higher excitations become progressively compressed and approach a common saturation region.}
    \label{fig:energy-levels}
\end{figure*}

\begin{figure*}[!htbp]
    \centering
    \subfigure[\(\ \)Charmonium summary bands.]{\includegraphics[width=0.47\textwidth]{charmonium_spectra_summary.pdf}}
    \subfigure[\(\ \)Bottomonium summary bands.]{\includegraphics[width=0.47\textwidth]{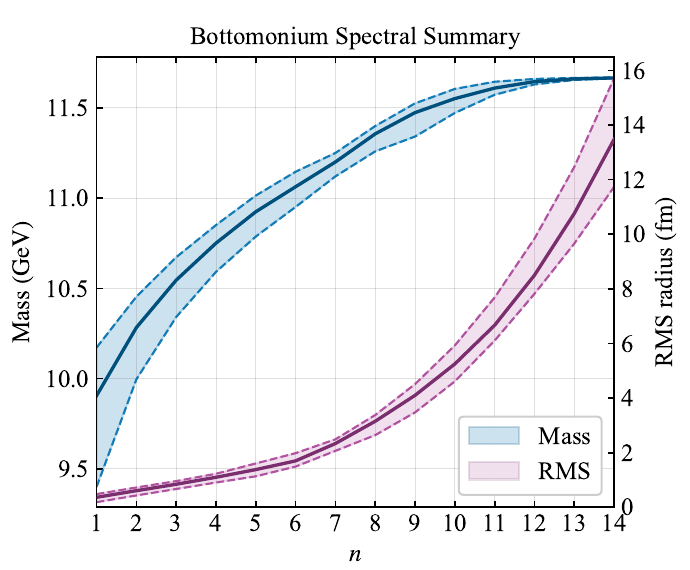}}
    \caption{Summary of the extended spectra in terms of the upper, lower, and average values across channels. 
    The mass bands show the convergence of the highly excited states toward a saturation region, 
    while the RMS radius bands display the rapid growth and divergence of the spatial size with increasing radial excitation.}
    \label{fig:spectra-summary}
\end{figure*}

The same pattern can be seen more directly in Fig.~\ref{fig:energy-levels}, which displays the calculated energy levels for the charmonium and bottomonium channels. 
In both systems, the low-lying levels remain well separated, while the higher excitations become increasingly compressed and accumulate near the limitation scale.
Fig.~\ref{fig:spectra-summary} presents the same information in a global form by showing the upper, lower, and average values across the channels. 
The narrowing of the mass band with increasing $n$ illustrates the convergence of the spectrum toward a common limiting region, 
while the widening of the RMS-radius band across channels indicates increasing variation of the highly excited states.
As shown in Figs.~\ref{fig:energy-levels} and \ref{fig:spectra-summary}, the high bottomonium excitations also begin to show numerical artifacts,
which are caused by double-precision round-off errors and the ill-conditioned eigenvalue problem. 
These results should therefore be regarded mainly as numerical references for the onset of the limitation region. 
Nevertheless, the presence and value of the limitation scale do not depend on the specific numerical implementation, 
but are controlled primarily by the confinement parameters $b$ and $\mu$.

The tabulated spectra also provide a useful reference for possible assignments of some observed states. 
In the bottomonium sector, the calculated spectrum leaves open the possibility of interpreting $\Upsilon(10753)$ as $\Upsilon(5S)$ and $\Upsilon(10860)$ as $\Upsilon_1(4D)$, 
because the predicted masses are close to the observed values. In the charmonium sector, the same comparison suggests that $\psi(4660)$ may be associated with the $\psi(7S)$ level, 
even though it has been regarded as $\psi(6S)$ in calculations with weaker screening effects and a smaller $\mu$ value. 
These assignments should be viewed as tentative within the present MGI framework, but they still provide a useful guide to highly excited quarkonium candidates.

\renewcommand{\tabcolsep}{0.18cm}
\renewcommand{\arraystretch}{1.1}
\begin{table*}[!htbp]
	\caption{Extended charmonium spectrum for the $nS$, $nP$, and $nD$ families with $n=1$--14. 
    For each channel, the calculated mass $M$ (in GeV) and root-mean-square radius $r$ (in fm) are listed.}
	\label{tab:charmonium-spectra}
	\begin{tabular}{cl|*{14}{c}}
		\toprule[1.5pt]\toprule[0.5pt]
		$^{2S+1}L_J$ &     & \multicolumn{14}{c}{$n$}                                                                                                                                                    \\
		             &     & 1      & 2      & 3      & 4      & 5      & 6      & 7      & 8      & 9      & 10     & 11     & 12     & 13     & 14     \\
		\midrule[0.5pt]
		$^1S_0$      & $M$ & 2.990 & 3.627 & 3.989 & 4.234 & 4.410 & 4.537 & 4.628 & 4.685 & 4.715 & 4.728 & 4.733 & 4.736 & 4.737 & 4.738 \\
		             & $r$ & 0.28 & 0.65 & 1.04 & 1.46 & 1.96 & 2.62 & 3.49 & 4.60 & 6.40 & 9.87 & 14.70 & 20.33 & 26.81 & 33.16 \\
		$^3S_1$      & $M$ & 3.103 & 3.672 & 4.015 & 4.252 & 4.422 & 4.546 & 4.634 & 4.690 & 4.718 & 4.729 & 4.734 & 4.736 & 4.737 & 4.738 \\
		             & $r$ & 0.32 & 0.69 & 1.07 & 1.50 & 2.00 & 2.69 & 3.64 & 4.93 & 6.97 & 10.67 & 15.57 & 21.37 & 28.56 & 35.07 \\
		$^1P_1$      & $M$ & 3.522 & 3.908 & 4.172 & 4.363 & 4.501 & 4.600 & 4.669 & 4.710 & 4.727 & 4.733 & 4.735 & 4.737 & 4.738 & 4.739 \\
		             & $r$ & 0.51 & 0.90 & 1.30 & 1.76 & 2.30 & 3.07 & 4.27 & 6.06 & 9.15 & 13.89 & 19.38 & 25.78 & 34.73 & 50.16 \\
		$^3P_0$      & $M$ & 3.419 & 3.851 & 4.135 & 4.337 & 4.483 & 4.589 & 4.663 & 4.707 & 4.726 & 4.732 & 4.735 & 4.737 & 4.738 & 4.739 \\
		             & $r$ & 0.45 & 0.83 & 1.24 & 1.69 & 2.23 & 3.00 & 4.16 & 5.87 & 8.73 & 13.31 & 18.71 & 25.13 & 34.19 & 48.97 \\
		$^3P_1$      & $M$ & 3.506 & 3.899 & 4.166 & 4.358 & 4.498 & 4.598 & 4.668 & 4.709 & 4.727 & 4.733 & 4.735 & 4.737 & 4.738 & 4.739 \\
		             & $r$ & 0.50 & 0.88 & 1.29 & 1.75 & 2.29 & 3.06 & 4.26 & 6.04 & 9.09 & 13.80 & 19.28 & 25.69 & 34.67 & 50.06 \\
		$^3P_2$      & $M$ & 3.563 & 3.935 & 4.191 & 4.377 & 4.511 & 4.607 & 4.673 & 4.711 & 4.727 & 4.733 & 4.735 & 4.737 & 4.738 & 4.739 \\
		             & $r$ & 0.55 & 0.93 & 1.34 & 1.81 & 2.35 & 3.11 & 4.31 & 6.14 & 9.33 & 14.14 & 19.68 & 26.05 & 34.85 & 50.39 \\
		$^1D_2$      & $M$ & 3.823 & 4.109 & 4.315 & 4.467 & 4.577 & 4.651 & 4.698 & 4.722 & 4.731 & 4.735 & 4.737 & 4.738 & 4.739 & 4.739 \\
		             & $r$ & 0.73 & 1.13 & 1.58 & 2.11 & 2.76 & 3.57 & 4.85 & 7.25 & 11.63 & 17.05 & 23.30 & 30.23 & 36.82 & 52.52 \\
		$^3D_1$      & $M$ & 3.787 & 4.083 & 4.296 & 4.452 & 4.566 & 4.644 & 4.694 & 4.721 & 4.731 & 4.734 & 4.736 & 4.738 & 4.739 & 4.739 \\
		             & $r$ & 0.68 & 1.09 & 1.52 & 2.04 & 2.65 & 3.45 & 4.75 & 7.04 & 11.26 & 16.62 & 22.75 & 29.44 & 36.87 & 53.08 \\
		$^3D_2$      & $M$ & 3.819 & 4.105 & 4.312 & 4.465 & 4.575 & 4.650 & 4.697 & 4.722 & 4.731 & 4.735 & 4.736 & 4.738 & 4.739 & 4.739 \\
		             & $r$ & 0.72 & 1.13 & 1.57 & 2.10 & 2.74 & 3.55 & 4.83 & 7.21 & 11.57 & 16.98 & 23.21 & 30.08 & 36.81 & 52.64 \\
		$^3D_3$      & $M$ & 3.843 & 4.123 & 4.326 & 4.475 & 4.583 & 4.656 & 4.701 & 4.723 & 4.731 & 4.735 & 4.737 & 4.738 & 4.739 & 4.739 \\
		             & $r$ & 0.76 & 1.16 & 1.61 & 2.15 & 2.84 & 3.68 & 4.97 & 7.46 & 11.95 & 17.41 & 23.77 & 31.01 & 37.00 & 51.74 \\
		\bottomrule[0.5pt]\bottomrule[1.5pt]
	\end{tabular}
\end{table*}

\renewcommand{\tabcolsep}{0.12cm}
\renewcommand{\arraystretch}{1.1}
\begin{table*}[!htbp]
	\caption{Extended bottomonium spectrum for the $nS$, $nP$, and $nD$ families with $n=1$--14. 
    For each channel, the calculated mass $M$ (in GeV) and root-mean-square radius $r$ (in fm) are listed.}
	\label{tab:bottomonium-spectra}
	\begin{tabular}{cl|*{14}{c}}
		\toprule[1.5pt]\toprule[0.5pt]
		$^{2S+1}L_J$ &     & \multicolumn{14}{c}{$n$}                                                                                                                                        \\
		             &     & 1                        & 2      & 3      & 4      & 5      & 6      & 7      & 8      & 9      & 10     & 11     & 12     & 13     & 14     \\
		\midrule[0.5pt]
		$^1S_0$      & $M$ & 9.400 & 9.996 & 10.341 & 10.591 & 10.787 & 10.951 & 11.121 & 11.259 & 11.341 & 11.472 & 11.573 & 11.630 & 11.655 & 11.664 \\
		             & $r$ & 0.19 & 0.43 & 0.67 & 0.91 & 1.13 & 1.49 & 2.07 & 2.64 & 3.48 & 4.61 & 6.11 & 8.02 & 10.44 & 13.66 \\
		$^3S_1$      & $M$ & 9.458 & 10.021 & 10.358 & 10.604 & 10.797 & 10.958 & 11.126 & 11.280 & 11.353 & 11.473 & 11.573 & 11.630 & 11.655 & 11.664 \\
		             & $r$ & 0.20 & 0.45 & 0.68 & 0.92 & 1.15 & 1.49 & 2.12 & 2.74 & 3.68 & 5.01 & 6.79 & 9.10 & 12.04 & 15.72 \\
		$^1P_1$      & $M$ & 9.892 & 10.259 & 10.524 & 10.733 & 10.906 & 11.042 & 11.191 & 11.357 & 11.491 & 11.540 & 11.593 & 11.639 & 11.658 & 11.665 \\
		             & $r$ & 0.34 & 0.58 & 0.82 & 1.07 & 1.30 & 1.56 & 2.35 & 3.29 & 4.18 & 5.12 & 6.19 & 7.83 & 9.68 & 11.77 \\
		$^3P_0$      & $M$ & 9.856 & 10.236 & 10.508 & 10.720 & 10.894 & 11.033 & 11.188 & 11.354 & 11.470 & 11.511 & 11.592 & 11.639 & 11.658 & 11.665 \\
		             & $r$ & 0.32 & 0.57 & 0.80 & 1.05 & 1.27 & 1.58 & 2.35 & 3.23 & 4.12 & 5.14 & 6.31 & 8.13 & 10.21 & 12.58 \\
		$^3P_1$      & $M$ & 9.886 & 10.256 & 10.522 & 10.731 & 10.905 & 11.041 & 11.191 & 11.357 & 11.489 & 11.534 & 11.592 & 11.639 & 11.658 & 11.665 \\
		             & $r$ & 0.34 & 0.58 & 0.82 & 1.07 & 1.30 & 1.56 & 2.35 & 3.28 & 4.05 & 4.99 & 6.19 & 7.82 & 9.66 & 11.74 \\
		$^3P_2$      & $M$ & 9.907 & 10.270 & 10.532 & 10.739 & 10.913 & 11.047 & 11.193 & 11.358 & 11.494 & 11.565 & 11.595 & 11.639 & 11.658 & 11.665 \\
		             & $r$ & 0.35 & 0.59 & 0.83 & 1.08 & 1.32 & 1.54 & 2.34 & 3.31 & 4.43 & 5.56 & 6.87 & 8.41 & 10.81 & 13.60 \\
		$^1D_2$      & $M$ & 10.165 & 10.449 & 10.668 & 10.847 & 11.011 & 11.141 & 11.249 & 11.400 & 11.525 & 11.605 & 11.645 & 11.661 & 11.666 & 11.669 \\
		             & $r$ & 0.48 & 0.72 & 0.95 & 1.22 & 1.59 & 1.95 & 2.43 & 3.25 & 4.27 & 5.55 & 7.12 & 9.03 & 11.35 & 14.15 \\
		$^3D_1$      & $M$ & 10.155 & 10.441 & 10.662 & 10.842 & 11.006 & 11.133 & 11.247 & 11.399 & 11.525 & 11.605 & 11.645 & 11.661 & 11.666 & 11.669 \\
		             & $r$ & 0.47 & 0.71 & 0.95 & 1.22 & 1.56 & 1.90 & 2.45 & 3.16 & 4.04 & 5.10 & 6.38 & 7.93 & 9.78 & 12.01 \\
		$^3D_2$      & $M$ & 10.164 & 10.448 & 10.668 & 10.847 & 11.011 & 11.140 & 11.249 & 11.400 & 11.525 & 11.605 & 11.645 & 11.661 & 11.666 & 11.669 \\
		             & $r$ & 0.48 & 0.72 & 0.95 & 1.22 & 1.58 & 1.95 & 2.43 & 3.24 & 4.25 & 5.50 & 7.04 & 8.91 & 11.17 & 13.91 \\
		$^3D_3$      & $M$ & 10.172 & 10.454 & 10.673 & 10.851 & 11.014 & 11.146 & 11.250 & 11.400 & 11.525 & 11.605 & 11.645 & 11.661 & 11.666 & 11.669 \\
		             & $r$ & 0.48 & 0.73 & 0.96 & 1.23 & 1.61 & 1.98 & 2.49 & 3.38 & 4.51 & 5.93 & 7.69 & 9.84 & 12.47 & 15.65 \\
		\bottomrule[0.5pt]\bottomrule[1.5pt]
	\end{tabular}
\end{table*}

\subsection{Limitation Analysis}

We now analyze the intrinsic limitation of the single-channel high radial excitations within the screened-potential framework.
The goal of this subsection is to identify the limitation scale of the spectrum and the corresponding limitation boundary at which physical interpretability is lost. 
The main criterion is the onset of mass-radius decoupling, namely the regime in which the mass spectrum becomes strongly compressed while the spatial size continues to grow rapidly.

First of all, the most important physical criterion follows from the asymptotic behavior of the screened confining interaction. In the large-$r$ region,
\begin{equation}
    \lim_{r\to\infty} \frac{b}{\mu}\left(1-e^{-\mu r}\right) = \frac{b}{\mu},
\end{equation}
so that the effective limitation scale tends to
\begin{equation}
    \lim_{r\to\infty} \left(m_1 + m_2 + \frac{b}{\mu}\left(1-e^{-\mu r}\right) + c\right)
    = m_1 + m_2 + \frac{b}{\mu} + c.
\end{equation}
The screened potential therefore produces a finite mass plateau, instead of an infinitely increasing spectrum deduced by a purely linear confining interaction.
The present analysis confirms that this plateau is located near 4.74 GeV for charmonium and near 11.67 GeV for bottomonium. 
Accordingly, the practical limitation scale is governed primarily by the confinement parameters $b$ and $\mu$, rather than by numerical implementation itself.

Second, the primary quantities used to define the limitation boundary are the adjacent mass gaps and corresponding RMS-radius increments,
\begin{equation}
    \Delta M_n = M_n - M_{n-1},
\end{equation}
\begin{equation}
    \Delta R_n = R_n - R_{n-1}.
\end{equation}
In practice, $\Delta M_n$ is expressed in MeV and $\Delta R_n$ in fm. The limitation boundary is defined through physical interpretability: 
a radial state is reliable only while the neighboring states remain both energetically resolvable and spatially compact. 
This condition begins to fail when the mass gaps become strongly compressed even though the RMS radii continue to grow rapidly. 
The decisive signature is therefore the onset of mass-radius decoupling. Because charmonium and bottomonium have different properties, 
the corresponding thresholds are chosen separately for the two systems.
For charmonium, the caution threshold is
\begin{equation}
    \min(\Delta M_n,\Delta M_{n+1}) < 10~\text{MeV},
    \qquad
    \max(\Delta R_n,\Delta R_{n+1}) > 3~\text{fm}.
\end{equation}
For bottomonium, the caution threshold is
\begin{equation}
    \min(\Delta M_n,\Delta M_{n+1}) < 10~\text{MeV},
    \qquad
    \max(\Delta R_n,\Delta R_{n+1}) > 2~\text{fm}.
\end{equation}
The analysis begins testing for limitation from $n=8$ onward. If $S=\{8,9,\dots,14\}$, the first caution state is defined by
\begin{equation}
    n_{\mathrm{caution}} = \min \left\{ n\in S : \min(\Delta M_n,\Delta M_{n+1}) < d_{\mathrm{caution}} \;\wedge\; \max(\Delta R_n,\Delta R_{n+1}) > r_{\mathrm{caution}} \right\}.
\end{equation}
The safe range ends at
\begin{equation}
    n_{\mathrm{safe}} = n_{\mathrm{caution}} - 1.
\end{equation}
Although these thresholds are empirical, they are physically well motivated and will be further tested by the additional diagnostics presented below.

\renewcommand{\tabcolsep}{0.4cm}
\renewcommand{\arraystretch}{1.1}
\begin{table*}[!htbp]
    \caption{Numerical summary of the limitation boundaries for the extended spectra. 
    The safe range ends at the state immediately before the first caution state. 
    The column $\Delta M_\textrm{caution}$ gives the mass gap in MeV, and $\Delta r_\textrm{caution}$ gives the RMS-radius jump in fm at the caution onset. 
    The next two columns list the changepoints inferred from the CUSUM and two-segment analyses of the indicator $f_n$. 
    The column $n_{\chi<1}$ gives the first index at which the dimensionless resolvability satisfies $\chi_n<1$, and $\chi_{\chi<1}$ is the corresponding value of $\chi_n$.}
    \label{tab:limitation-summary}
    \begin{tabular}{llcccccccc}
        \toprule[1.5pt]
        & & \multicolumn{4}{c}{Threshold-based} & \multicolumn{2}{c}{Changepoint-based} & \multicolumn{2}{c}{$\chi$-based} \\
        \cmidrule[0.5pt](lr){3-6} \cmidrule[0.5pt](lr){7-8} \cmidrule[0.5pt](lr){9-10}
        \multicolumn{2}{l}{Channel} & $n_\textrm{safe}$ & $n_\textrm{caution}$ & $\Delta M_\textrm{caution}$ & $\Delta r_\textrm{caution}$ 
        & $n_\textrm{CUSUM}$ & $n_\textrm{2-seg}$ & $n_{\chi<1}$ & $\chi_{\chi<1}$ \\
        \midrule[0.5pt]
         $b\bar{b}$ & $^1S_0$ & 12 & 13 & 8.99 & 3.23 & 8 & 9 & 14 & 0.623 \\
                    & $^3S_1$ & 12 & 13 & 8.99 & 3.68 & 8 & 9 & 14 & 0.716 \\
                    & $^1P_1$ & 12 & 13 & 6.91 & 2.09 & 9 & 10 & 13 & 0.954 \\
                    & $^3P_0$ & 12 & 13 & 6.93 & 2.38 & 8 & 10 & 14 & 0.442 \\
                    & $^3P_1$ & 12 & 13 & 6.92 & 2.08 & 9 & 10 & 13 & 0.922 \\
                    & $^3P_2$ & 12 & 13 & 6.90 & 2.79 & 8 & 10 & 14 & 0.475 \\
                    & $^1D_2$ & 11 & 12 & 5.46 & 2.32 & 9 & 11 & 12 & 0.709 \\
                    & $^3D_1$ & 12 & 13 & 2.63 & 2.22 & 9 & 11 & 12 & 0.623 \\
                    & $^3D_2$ & 11 & 12 & 5.46 & 2.27 & 9 & 11 & 12 & 0.710 \\
                    & $^3D_3$ & 11 & 12 & 5.46 & 2.62 & 9 & 11 & 12 & 0.785 \\
         $c\bar{c}$ & $^1S_0$ & 9 & 10 & 4.81 & 4.83 & 8 & 10 & 9 & 0.993 \\
                    & $^3S_1$ & 9 & 10 & 4.22 & 4.90 & 8 & 10 & 10 & 0.599 \\
                    & $^1P_1$ & 8 & 9 & 5.74 & 4.74 & 8 & 9 & 9 & 0.988 \\
                    & $^3P_0$ & 8 & 9 & 6.39 & 4.58 & 8 & 9 & 9 & 0.990 \\
                    & $^3P_1$ & 8 & 9 & 5.83 & 4.72 & 8 & 9 & 9 & 0.991 \\
                    & $^3P_2$ & 8 & 9 & 5.48 & 4.81 & 8 & 9 & 9 & 0.979 \\
                    & $^1D_2$ & 7 & 8 & 8.90 & 4.38 & 8 & 9 & 8 & 0.988 \\
                    & $^3D_1$ & 7 & 8 & 9.87 & 4.22 & 8 & 9 & 8 & 0.996 \\
                    & $^3D_2$ & 7 & 8 & 9.05 & 4.36 & 8 & 9 & 8 & 0.994 \\
                    & $^3D_3$ & 7 & 8 & 8.17 & 4.49 & 8 & 9 & 8 & 0.990 \\
        \bottomrule[1.5pt]
    \end{tabular}
\end{table*}

The channel-by-channel limitation boundaries are summarized in Table~\ref{tab:limitation-summary}. 
The results exhibit a clear and consistent pattern across channels. Charmonium reaches the limitation region earlier than bottomonium: 
the $D$-wave channels are the first to enter the caution region at $n=8$, followed by the $P$-wave channels at $n=9$ and the $S$-wave channels at $n=10$. 
In bottomonium, the same transition appears later, with the $D$-wave channels entering caution at $n=12$ and the $S$- and $P$-wave channels at $n=13$. 
Consistently, the practical safe ranges are about $n=1$--7 to $1$--9 for charmonium and $n=1$--11 to $1$--12 for bottomonium. 
At these limitation boundaries, the mass gaps have already compressed to a few MeV, while the RMS-radius jumps remain sizable, at the level of several fm.
This combination is the signal of mass-radius decoupling and therefore denotes the onset of the spectrum limitation region.

Next, a supporting changepoint test may be constructed entirely from spectrum data. A convenient choice is to define a monotonic convergence/divergence indicator
\begin{equation}
    f_n = \log_{10}\!\left(\frac{\Delta r_n + \epsilon_r}{\Delta M_n + \epsilon_M}\right),
\end{equation}
where $\epsilon_r=0.001\,\mathrm{fm}$ and $\epsilon_M=0.001\,\mathrm{MeV}$ are small positive values introduced only to avoid numerical instability. 
The quantity $f_n$ increases when the RMS radius grows while the mass compresses, and therefore provides a convenient one-dimensional summary of the spectral transition.
Two standard changepoint procedures may be applied to the sequence $\{f_n\}$. First, one defines the standardized variable
\begin{equation}
    z_n = \frac{f_n - \bar f}{\sigma_f},
\end{equation}
where $\bar f$ and $\sigma_f$ are the mean and standard deviation of the sequence, then constructs the cumulative-sum (CUSUM) process
\begin{equation}
    S_k = \sum_{n \le k} z_n.
\end{equation}
If the sequence is statistically homogeneous, the positive and negative fluctuations tend to offset each other, so $S_k$ remains relatively balanced. 
If the sequence contains a transition from an early regime to a later one, $S_k$ begins to show a systematic drift. The corresponding changepoint is taken as
\begin{equation}
    n_{\mathrm{CUSUM}} = \operatorname*{arg\ max}_k |S_k|.
\end{equation}
As a second supporting test, the sequence is split at a candidate index $j$ into early and late segments, and the residual sum of squares is defined by
\begin{equation}
    \mathrm{RSS}(j) = \sum_{n < j} \left(f_n - \bar f_{<j}\right)^2 + \sum_{n \ge j} \left(f_n - \bar f_{\ge j}\right)^2,
\end{equation}
where $\bar f_{<j}$ and $\bar f_{\ge j}$ are the means of the left and right segments, respectively. The two-segment changepoint is then
\begin{equation}
    n_{\mathrm{2-seg}} = \operatorname*{arg\ min}_j \mathrm{RSS}(j).
\end{equation}
In practice, CUSUM detects the point of largest cumulative drift, while the two-segment method identifies the split that gives the best two-regime description of the sequence.
The changepoint results included in Table~\ref{tab:limitation-summary} are qualitatively consistent with the primary limitation boundaries. 
In particular, both methods place the transition systematically earlier in charmonium than in bottomonium. 
This supports the limitation boundary being associated with a structural transition in the spectrum,
while the final conclusion is still determined by the primary physical criterion.

Finally, an additional supporting quantity is the dimensionless combination
\begin{equation}
    \chi_n = \Delta M_n r_n,
\end{equation}
where $\Delta M_n$ is expressed in GeV and $r_n$ in $\mathrm{GeV}^{-1}$. 
This quantity provides a compact measure of resolvability by combining the neighboring mass gap and the state size into a dimensionless number. 
The corresponding results are included in Table~\ref{tab:limitation-summary} through the columns $n_{\chi<1}$ and $\chi_{\chi<1}$. 
They show that $\chi_n$ drops below unity much earlier in charmonium, typically around $n=8$--10, than in bottomonium, where this happens mainly around $n=12$--14. 
These $\chi$-based results are consistent with the previous threshold-based analysis. 
Therefore, the behavior of $\chi_n$ provides an additional and independent support for the limitation boundary analysis.

In summary, the limitation boundary is identified primarily by the physical mass-radius decoupling criterion, 
and is further tested by the changepoint analysis and the dimensionless resolvability measure $\chi_n$. 
All of these diagnostics lead to the same qualitative conclusion: charmonium reaches the limitation region earlier than bottomonium, 
and the high-lying states beyond the safe range are no longer robustly identifiable as compact quarkonium states. 
More specifically, the limitation scale is about $4.74\,\mathrm{GeV}$ for charmonium and $11.67\,\mathrm{GeV}$ for bottomonium, 
while the corresponding limitation boundaries appear around $n=8$--10 and $n=12$--13, respectively. 
In addition to this physical criterion, the RMS radii at the limitation boundary are already around $10\,\mathrm{fm}$, 
which is about one order of magnitude larger than the QCD confinement scale of $\sim 1\,\mathrm{fm}$. 
This further supports the validity of the criterion and the resulting conclusions.